\documentstyle[eqsecnum,aps,psfig]{revtex}
\begin{document}
\title{Thermodynamics of the 2D Hubbard model}
\author{F. Mancini\footnote{Corresponding author.  E-mail address:
mancini@vaxsa.csied.unisa.it}, D. Villani}
\address{Dipartimento di Scienze Fisiche ``E.R. Caianiello" e Unit\`a
I.N.F.M. di Salerno \\ Universit\`a di Salerno, 84081 Baronissi (SA),
Italy}
\author{H. Matsumoto}
\address{Department of Applied Physics, Seikei University, Tokyo 180,
Japan}
\maketitle
\begin{abstract}
 A theoretical analysis of the thermodynamic response functions of the
2D single-band Hubbard model is carried out by means of the composite
operator method. It is shown that all the features of these quantities
can be explained by looking at the dependence of the thermodynamic
variables on their conjugate ones, that is, for example, the relation
between entropy and temperature, chemical potential and particle
concentration, double occupancy and on-site Coulomb repulsion. In this
way, the electronic specific heat and the entropy per site are
determined in the paramagnetic phase. Also, for the electronic specific
heat and internal energy we present three different schemes of
calculation. A comprehensive comparison among them for the interacting
case opens the possibility to obtain a deep theoretical understanding of
the computed quantities whereas, for the non-interacting case, only
confirms the internal consistency of the thermodynamic approach. It is
found that the numerical data from quantum Monte Carlo techniques for
the internal energy and electronic specific heat are well reproduced by
determining them through the first and second temperature derivatives of
the chemical potential. The anomalous normal state properties in
hole-doped cuprate high $T_c$ superconductors are also well described.
Actually, the results for the linear coefficient of the electronic
specific heat are in agreement with those obtained by using a pure
fermionic theoretical scheme. Indeed, the results for the Wilson ratio
seem to confirm the same scenario. Finally, we obtain several
characteristic crossing points for the response functions when plotted
versus some thermodynamic variables. These peculiar features indicate
the existence of more than one energy scale competing with thermal
excitations and indicate, as already noted by Volhardt, a crossover from
a non-interacting to a highly correlated behavior.
\end{abstract}

\section{INTRODUCTION}

 It is believed, on both experimental and theoretical grounds, that
superconductivity and charge transport in high $T_c$ cuprates are mostly
confined to the $CuO_2$ planes [1,2]; and hence the attention of many
physicists has been dedicated to 2D models which contain as an essential
feature a competition between the band picture and highly correlated
many body effects. Of course, some features of the phase diagram, like
the existence of a finite N\'eel temperature, can only be explained by
adding a coupling between the planes.

 The bonding combination of $Cu$ and $O$ orbitals turns out to be quite
deep below the Fermi level, so that no dynamical freedom is left to
treat $d$  and $p$  orbitals separately [3] (there are some strong
experimental evidences, mostly based on the study of the Knight shift,
that in the $CuO_2$ plane one spin degree of freedom is observed [4]).
Through the Pauli principle, the energy of the $p$ electron excitation
is, for example, largely modified by the change of charge and spin
states of the neighboring $Cu$ ions. A $p$ electron and charge and spin
fluctuations on neighboring $Cu$ ions are simultaneously excited so that
electronic excitations are formed on a $CuO_2$ cluster as a whole. Then,
the resulting complex can be described by a single-band Hubbard model
[5].

 In the simplest form, the Hubbard model, first introduced to describe
the correlations of electrons in a narrow $d$-band of transition metals,
contains a kinetic term which describes the motion of the electrons
among the sites of the Bravais lattice and an interaction term between
electrons of opposite spin on the same lattice site. By varying the
model parameters, it is believed that the Hubbard model is capable of
describing many properties of strongly correlated fermion systems. Among
different examples, the Hubbard Hamiltonian is applicable to describe
the metal-insulator transition in a series of transition metal oxides
such as $Sr_{1-x} La_x Ti O_3$ [6,7] and $V_2 O_3$ [8-11]. The
applicability of the model to the superconducting copper-oxides is
related to the fact that upon doping most of these compounds exhibit a
metal-insulator Mott transition; the superconducting state is near the
N\'eel state and there are many experimental results [12-15] which show
a close relation between the antiferromagnetic correlations in the
$Cu-O$ planes and the occurrence of the superconducting phase. However,
it is important to stress that an appropriate description of a bad metal
with large energy scale spin fluctuations by means of a purely
electrostatic Hamiltonian should preserve the symmetry expressed by the
Pauli principle that codifies the correct interplay between charge and
magnetic configurations [16-18].

 Although considerable attention has been devoted to the Hubbard model
and significant progress was achieved in understanding ground state
properties, particularly at half-filling, static and dynamical spin
correlations, the optical conductivity and other observables, a clear
comprehension of the low-lying excitations is still lacking [2]. The
difficulty is not to be found only in the absence of any obvious small
parameter in the strong coupling regime. More deeply, it is due to the
difficulty of handling simultaneously itinerant aspects (spatial
correlations) and atomic aspects (pronounced on- site quantum
fluctuations) [19].

 In recent years we have been developing a method of calculation,
denominated Composite Operator Method (COM) [17,18,20-32], that has been
revealed to be a powerful tool for the description of local and
itinerant excitations in strongly correlated systems. In previous
papers, we considered the Hubbard electronic operators for the
determination of fundamental excitations. A fully self-consistent
calculation of the electronic propagator has been realized by means of a
constraint with the physical content of the Pauli principle [17, 23-25].
We calculated local quantities, as the double occupancy and the magnetic
moment [17,23], the energy per site [24], the chemical potential [24],
the magnetic susceptibility [25, 29], the density of states and the
quasi-particle spectra [28, 32]. In all the cases, the results show a
good agreement with those obtained by numerical simulation. In
particular, the results obtained for the magnetic properties can
reproduce the unusual characteristics observed in high $T_c$
superconducting materials [25, 29, 33]. Therefore, the agreement
strengthens the idea that a microscopic single-band model contains the
essential physical features of the new class of materials.

 In this paper we investigate the electronic specific heat and the
entropy per site of the 2D Hubbard model for a paramagnetic ground
state. It will be shown that all the features of these quantities can be
understood by looking at the dependence of the chemical potential and
double occupancy on their conjugate thermodynamic variables, that is,
the particle concentration and the on- site Coulomb repulsion,
respectively. A comprehensive comparison among different methods to
compute the specific heat will shed a new light on the approximation
used. It will emerge that in our theoretical scheme, even if dynamical
effects in the self-energy are neglected promoting unstable collective
asymptotic modes to the role of well-defined quasi-particle excitations,
extended spin modes can be captured by properly combining symmetry
requirements and extended operatorial basis. Indeed, the presence of a
low temperature peak that appears when the low-lying spin states are
excited will appear as an important feature shared with the quantum
Monte Carlo data [34]. In addition, an alternative fermionic scenario
for the computation of the electronic specific heat together with the
results for the Wilson ratio will confirm cuprates as dominated by
conventional fermionic excitations. An extensive study of the
thermodynamic response functions will reveal the existence of critical
lines which separate different energy scales created by the interplay
between charge and spin modes. In other words a study of thermodynamics
quantities, such as the double occupancy, the entropy, the chemical
potential, the specific heat, indicates lines in the $U-T$ plane which
separate a highly-correlated behavior, dominated by spin and charge
fluctuations and a non-interacting behavior, dominated by thermal
fluctuations. In particular, there emerges a region of filling where the
entropy reduces by increasing the filling signalling the set up of an
ordered phase. For $T\to 0$ there is a well-defined marginal
concentration where a quantum phase transition occurs. A detailed
comparison with the non-interacting case will be also presented
throughout the paper.

 The plan of the article is as follows. In the next Section we present
the 2D Hubbard model and the electron propagator in the COM. In Sec. III
we review experimental data for some thermodynamic properties. The
results for the electronic specific heat are presented in Sec IV, where
a theoretical understanding of the different ways to compute the
specific heat is also presented. Section V is devoted to a discussion of
double occupancy. In Sec. VI the results for the chemical potential
versus temperature are discussed. The entropy is analyzed in Sec VII.
Some concluding remarks are presented at the end.

\section{ELECTRON PROPAGATOR IN THE HUBBARD MODEL}

 The Hubbard model is defined by
\begin{equation}
H=\sum_{i,j} t_{ij} c^\dagger (i) \cdot c(j) + U \sum_i n_\uparrow (i)
n_\downarrow (i) -\mu\sum_i n(i)
\end{equation}            
The notation is the following. The variable $i$  stands for the lattice
vector ${\rm \bf R} _i$. $\left\{ c(i), c^\dagger(i) \right\}$ are
annihilation and creation operators of $c-$electrons at site $i$, in the
spinor notation:
\begin{equation}
c=\left({c_\uparrow \atop c_\downarrow}\right)\qquad c^\dagger=\left(
c^\dagger_\uparrow \quad c^\dagger_\downarrow  \right)
\end{equation}          
$t_{ij}$ denotes the transfer integral and describes hopping between
different sites; the U term is the Hubbard interaction between two
$c-$electrons at the same site with
\begin{equation}
n_\sigma(i) = c^\dagger_\sigma(i) c_\sigma(i)
\end{equation}     
being the charge-density operator per spin $\sigma$. $n(i)$ is the total
charge-density operator.  $\mu$ is the chemical potential. In the
nearest neighbor approximation, for a two-dimensional cubic lattice with
lattice constant $a$, we write the hopping matrix $t_{ij}$ as
\begin{equation}
t_{ij}=-4t\alpha_{ij} = -4t{1\over N}\sum _{\rm \bf k} e^{i{\rm \bf
k}\cdot (\rm
\bf r_i-{\rm \bf R}_j)} \alpha ({\rm \bf k})
\end{equation}             
where
\begin{equation}
\alpha({\rm \bf k}) = {1\over 2} \left[ \cos ( k_xa) + \cos (k_ya)\right]
\end{equation}             
The scale of the energy has been fixed in such a way that  $t_{ii}=0$.
It should be noted that since the interactions are restricted to the
same site, the dimensionality of the system comes in only when a
specific form for  $\alpha({\rm \bf k})$ is taken. In other words, the
stabilization of eventual cooperative phenomena is uniquely governed by
the band dispersion.

 The point of view adopted in the COM is that the Heisenberg operators
$\{ c(i), c^\dagger(i)\}$ are not good candidates as a basis for
calculations. Because of strong correlations the $c-$electrons loose
their identity and new fields, whose properties are self-consistently
determined by the dynamics and by the symmetries of the model, together
with the boundary conditions, might be more appropriate as a starting
point for the physical description of the system. Due to the on-site
Coulomb interaction, it is known that two sharp features develop in the
band structure which correspond to the Hubbard subbands and describe
interatomic excitations mainly restricted to subsets of the occupancy
number. Indeed, a first natural choice for composite fields is given by
the Hubbard constrained electronic operators
\begin{eqnarray}
\xi(i) &=& [1-n(i)] c(i)\\   
\eta(i) &=& n(i)c(i)    
\end{eqnarray}
describing the transitions $(n=0)\Longleftrightarrow (n=1)$ and
$(n=1)\Longleftrightarrow (n=2)$, respectively. The two-point retarded
thermal Green's function is defined as
\begin{equation}
S(i,j)=\langle R[\Psi(i) \Psi^\dagger (j)]\rangle\ ,
\end{equation}                           
where $\Psi(i)$ is the doublet composite operator
\begin{equation}
\Psi(i)={\xi(i) \choose \eta(i)}
\end{equation}           
The bracket $\langle \ldots \rangle$   indicates the thermal average and
$R$  is the usual retarded operator.

In previous papers, we have shown that the determination of the
single-particle Green's function (2.8) can be realized in a fully
self-consistent way once a unique approximation is made [17, 23-25].
This approximation consists in neglecting the dynamical part in the
self-energy and corresponds to a pole expansion of the spectral
intensities. As shown in Ref. 23-25, by considering time translational
invariance and no magnetic order, the single-particle electronic
propagator has the following expression
\begin{eqnarray}
S_{cc} (i,j) &=& \langle R[ c (i) c^\dagger (j)] \rangle = \nonumber \\
&=& {i\Omega\over (2\pi)^3} \int_{\Omega_B} d^2 kd\omega
e^{i{\rm \bf k}\cdot({\rm \bf R}_i-{\rm \bf R}_j)-i\omega(t_i-t_j)}
[1-f_F(\omega)]
\left[{A_1({\rm \bf k})\over \omega- E_1({\rm \bf k}) +i\eta} + {A_2({\rm \bf k})\over \omega-
E_2({\rm \bf k}) +i\eta}
\right]
\end{eqnarray}                           
$\Omega$ and $\Omega_B$ being the volume of the unit cell in the direct
and reciprocal space, respectively. $f_F(\omega)$ is the Fermi
distribution function. $E_1({\rm \bf k})$ and $E_2({\rm \bf k})$ are the
energy spectra
\begin{equation}
E_{1,2}({\rm \bf k})=R({\rm \bf k}) \pm Q({\rm \bf k})
\end{equation}              
$A_1({\rm \bf k})$ and $A_2({\rm \bf k})$ are the spectral intensities
\begin{eqnarray}
A_1({\rm \bf k}) =& \displaystyle {1\over 2} \left[1-{U(1-n)\over
2Q({\rm \bf k})}+{m_{12} ({\rm \bf k})\over 2I_{11}I_{22} Q({\rm \bf
k})}\right]
\\
\\
A_2({\rm \bf k}) =&  \displaystyle {1\over 2} \left[1+{U(1-n)\over
2Q({\rm \bf k})}-{m_{12} ({\rm \bf k})\over 2I_{11}I_{22} Q({\rm \bf
k})}\right]
\end{eqnarray}       
with the definitions
\begin{eqnarray}
R &=& \displaystyle {1\over 2} U - \mu - {2t\over I_{11} I_{22}} \left\{
\Delta +\alpha ({\rm \bf k}) [p+(1-n) I_{22}]\right\} \\
\\
Q &=& \displaystyle {1\over 2} \sqrt{U^2 +{1\over I_{11}^2 I_{22} ^2}
m_{12}^2 - 2U{(1-n)\over I_{11}I_{22}} m_{12}}\\
\\
m_{12}({\rm \bf k}) &=& 4t\left[\Delta +\alpha({\rm \bf k})
\left( p-{n\over 2}\right)\right]
\end{eqnarray}
where  $n=\langle c^\dagger c\rangle$ is the particle density. The
parameters $\Delta$ and $p$ are static intersite correlation functions
defined as
\begin{equation}
\Delta\equiv \langle \xi^\alpha(i) \xi^\dagger(i)\rangle - \langle
\eta^\alpha(i) \eta^\dagger (i)\rangle
\end{equation}       
\begin{equation}
p\equiv {1\over 4} \langle n^\alpha_\mu (i) n_\mu(i) \rangle - \langle
[c_\uparrow(i) c_\downarrow (i) ]^\alpha c^\dagger_\downarrow(i)
c^\dagger_\uparrow(i)\rangle
\end{equation}                    
The notation $c^\alpha(i)$ stands to indicate the field $c$ on the first
neighbor sites:
\begin{equation}
c^\alpha(i)=\sum_j \alpha_{ij} c(j)
\end{equation}         
We are also using the following notation
\begin{equation}
I_{11}= 1-n/2\qquad I_{22} = n/2
\end{equation}           

 A crucial point in the calculation of the propagator (2.10) is the
implementation of the Pauli principle by means of constraining equations
to determine the internal parameters $\mu$, $\Delta$ and $p$. Use of
this principle leads to the self-consistent equations [23, 25]
$$
n=1-G_0 + U(1-n) F_0
$$
\begin{equation}
\Delta = {1-n\over 2}G_1 - {U\over 2}F_1 + {(1-n)\over 2I_{11} I_{22}
}B_1
\end{equation}               
$$
pF_1 = I_{22} F_1 - \Delta F_0
$$
where
$$
F_n = {\Omega\over 2(2\pi)^2}\int _{\Omega_B} d^2 k [\alpha ({\rm \bf
k})]^n f({\rm \bf k})
$$
\begin{equation}
G_n = {\Omega\over 2(2\pi)^2}\int _{\Omega_B} d^2 k [\alpha ({\rm \bf
k})]^n g({\rm \bf k})
\end{equation}                  
$$
B_n = {\Omega\over 2(2\pi)^2}\int _{\Omega_B} d^2 k [\alpha ({\rm \bf
k})]^n f({\rm \bf k})m_{12}({\rm \bf k})
$$
with
\begin{eqnarray}
 f({\rm \bf k}) =\displaystyle  {T_1({\rm \bf k}) - T_2({\rm \bf
 k})\over 2Q({\rm \bf k})} \qquad g({\rm \bf k})
=[T_1({\rm \bf k})+T_2({\rm \bf k})]\\
\\
T_i({\rm \bf k}) = \displaystyle \tanh\left({E_i({\rm \bf k})\over
2k_BT}\right)
\end{eqnarray}         
The recovery of the Pauli principle, very often violated by other
approximations, assures a dynamics bounded to the Hilbert space capable
of describing in a correct way the interplay between the charge and the
magnetic configurations. Furthermore, we have shown [35] that in the
two-pole approximation [36] the set of self-consistent equations (2.18)
is the only one which restores the particle-hole symmetry and the Pauli
principle, which are intimately connected.

 It is possible to go beyond the two-pole approximation by enlarging the
set of asymptotic fields [18, 20-22] or by taking into account the
dynamical corrections to the self-energy [31,32].

\section{THERMODYNAMICS AS REVEALED BY EXPERIMENTS}

 The electronic specific heat $C(T)$ of cuprate high-$T_c$
superconductors has been measured. In particular $C(T)$ of $La_{2-x}
Sr_x CuO_4$ [37, 38] has been studied for $0.03<x<0.45$ in the range of
temperatures between 1.5 and 300 K, and of $YBa_2Cu_3 O_{6+y}$ [39] for
$0.16\le y\le 0.97$ between 1.8 and 300 K. From these experiments the
following behavior has been observed for the coefficient $\gamma=C/T$ of
the normal state specific heat:

\noindent a) for fixed temperature, $\gamma(x,T)$ increases with doping;

\noindent a1) in the case of $La_{2-x} Sr_x CuO_4$, $\gamma(x,T)$
exhibits a rather sharp maximum at $x\approx 0.25$  (near the doping
where superconductivity disappears), then starts to decrease; the same
behavior for  $\gamma(x,T)$ has been estimated in Ref. 40, but with a
peak located around $x\approx 0.18$, close to the optimal doping; for
$La_{2-x} Ba_x CuO_4$ [41] a maximum has been observed at
$x\approx0.22$;

\noindent a2) in the case of $YBa_2Cu_3O_{6+y}$, $\gamma(x,T)$ increases
smoothly to a plateau or two broad maxima, situated at $y\approx 0.6$
and $y\approx 0.9$, respectively;

\noindent b) for fixed doping, $\gamma(x,T)$ as a function of
temperature exhibits a broad peak moving to lower temperatures with
increasing the dopant concentration;

\noindent c) further increasing $y$, the T-dependence weakens and in the
region of high doping no increase is observed. For $YBa_2Cu_3O_{6+y}$,
no substantial increase is observed for $y>0.8$.

 As noticed by Vollhardt [42], there is a peculiar feature of the
specific heat observed in a large variety of systems. The specific heat
curves versus T, when plotted for different, not too large, values of
some thermodynamic variable, intersect at one or even two well defined
temperatures. In  $^3He$ the specific heat $C(T,P)$ curves versus T at
different pressures P intersect at a well defined temperature [43,44];
in heavy fermions $CeAl_3$ [45] and  $UBe_{13}$ [46] upon change of $P$,
$UPt_{3-x} Pd_x$ [47] and $CePt_3 Si_{1-x} Ge_x$ [48] upon change of
$x$, $CeCu_{6-x} Au_x$  when either P [49] or the magnetic field B [50]
is varied; in semi-metal, $Eu_{0.5}Sr_{0.5}As_3$ [51] upon change of B.

 The following properties have been observed for the entropy $S$  [38,
39, 52]:

\noindent a) for a given temperature, $S$ increases with doping;

\noindent a1) in the case of $La_{2-x} Sr_x CuO_4$ [38], $S(x,T)$
reaches a maximum in the vicinity of $x\approx 0.25$, then decreases;

\noindent a2) in the case of $YBa_2Cu_3O_{6+y}$ [39, 52], $S$ reaches a
maximum in the vicinity of $y\approx 0.97$;

\noindent b) for a given dopant concentration, $S$ exhibits a
superlinear dependence on the temperature;

\noindent c) the normal state entropy as a function of  $T$ extrapolates
to a negative value at $T=0K$;

\noindent d) there is a striking numerical correlation between $S/T$ and
$a\chi_0$, where $\chi_0$ is the bulk susceptibility and  $a$ is the
Wilson ratio.

\section{ELECTRONIC SPECIFIC HEAT}
\subsection{GENERAL FORMULAS}
The specific heat $C(T)$ is defined as
\begin{equation}
C(T)={dE\over dT}
\end{equation}        
where  $E$  is the internal energy density, given by the thermal average
of the Hamiltonian
\begin{equation}
E = {1\over N} \langle H\rangle
\end{equation}       
$N$ being the number of sites. Calculation of internal energy by means
of Eq. (4.2) will generally require the calculation of two-particle
Green's functions. An alternative way to calculate the internal energy
is the following. By introducing the Helmholtz free energy per site
\begin{equation}
F= E-TS
\end{equation}    
where $S$ is the entropy per site, from the thermodynamics we have
\begin{equation}
S=-\left({\partial F \over\partial T}\right)_n \qquad
\mu=\left({\partial F \over\partial n}\right)_T \qquad
\left({\partial S \over\partial n}\right)_T =-\left({\partial \mu
\over\partial T}\right)_n
\end{equation}            
Then, it is straightforward to obtain the following formulas
\begin{equation}
F(T,n)=\int_0^n \mu (T, n') dn'
\end{equation}            
\begin{equation}
S(T,n)=-\int_0^n \left({\partial \mu \over\partial T}\right)_{n'} dn'
\end{equation}       
\begin{equation}
E(T,n)=\int_0^n\left[ \mu(T, n') - T\left({\partial \mu \over\partial
T}\right)_{n'}\right]dn'
\end{equation}               
from which the specific heat turns out to be
\begin{equation}
C(T,n) =-T\int_0^n \left({\partial^2 \mu \over\partial T^2}\right)_{n'}
dn'
\end{equation}       
In this scheme the thermodynamic quantities are all expressed through
the chemical potential, whose determination requires  the knowledge of
the single-particle Green's function.

 Summarizing, we have two distinct ways to calculate the internal
energy, based on the use of Eqs. (4.2) and (4.7). In principle these
equations are equivalent and lead to the same result when an exact
solution is available. However, the situation drastically changes when
approximations are involved and different results can be obtained.
Indeed an open problem in Condensed Matter Physics is to find a unique
consistent scheme of approximation capable of treating on an equal
footing, both one- and two-particle Green's functions.

Sometimes in the literature the internal energy for electronic systems
is calculated by means of the expression
\begin{equation}
E=\int_{-\infty}^{+\infty}d\omega N(\omega)f_F (\omega)\omega
\end{equation}                 
where $N(\omega)$ is the density of states. However, Eq (4.9) is based
on the assumption that the system admits a description in terms of
fermionic quasi-particles. It is well known that the correlation among
the original electrons can generate extended bosonic modes, and use of
expression (4.9) requires special attention. By general argument it can
be shown that for interacting systems the correct expression for the
internal energy, calculated by means of the density of states, is given
by
\begin{equation}
E=\int_{-\infty}^{+\infty}d\omega N(\omega)f_F (\omega)\omega -
\langle H_I\rangle
\end{equation}
where $\langle H_I\rangle$ is the non-quadratic part of the Hamiltonian
in the chosen canonical representation.

\subsection{NON-INTERACTING CASE}

 To discuss the specific heat it is useful at first to consider the
non-interacting [i.e. U=0] Hubbard model. In this case the thermal
retarded Green's function can be exactly calculated and has the
expression
\begin{eqnarray}
S_{cc}(i,j)&=&\langle R[c(i)c^\dagger (j)]\rangle \nonumber \\
&=& {i\Omega\over (2\pi)^3} \int_{\Omega_B}
d^2 kd\omega e^{i{\rm \bf k}\cdot ({\rm \bf R}_i-{\rm \bf
R}_j)-i\omega(t_i-t_j)} [1-f_F(\omega)] {1\over \omega-E({\rm \bf k})
+i\eta}
\end{eqnarray}          
where the energy spectrum has the expression
\begin{equation}
E({\rm \bf k}) = -\mu-4t\alpha({\rm \bf k})
\end{equation}        
The chemical potential is determined as a function of n and T by means
of the equation
\begin{equation}
n={2\Omega\over (2\pi)^2} \int_{\Omega_B} d^2 kf_F [E({\rm \bf
k})]=1-{\Omega\over (2\pi)^2}
\int_{\Omega_B} d^2 k T({\rm \bf k})
\end{equation}            
where we put
\begin{equation}
T({\rm \bf k}) = \tanh\left({E({\rm \bf k})\over 2k_BT}\right)
\end{equation}       
As we discussed above, we have different ways of calculating the free
energy. In the non-interacting case, where an exact solution is
available, all different procedures must give the same result. Since
this point will acquire some relevance in the interacting case, we shall
examine this in detail.

 In the non-interacting case the density of states is given by
\begin{equation}
N(\omega) = {\Omega\over (2\pi)^2} \int_{\Omega_B} d^2 k \delta[\omega -
E({\rm \bf k})]
\end{equation}                
By substituting this expression into Eq. (4.10) we have
\begin{equation}
E={4t\Omega\over (2\pi)^2} \int_{\Omega_B}d^2 k\alpha({\rm \bf k})
T({\rm \bf k})
\end{equation}       
where a factor 2 has been included in order to take into account the
spin degree of freedom.

The same result is obtained by taking the thermal average of the
Hamiltonian
\begin{equation}
E=8t\langle c^\alpha (i) c^\dagger(i) \rangle={4t\Omega\over (2\pi )^2}
\int_{\Omega_B}
d^2 k\alpha ({\rm \bf k})T({\rm \bf k})
\end{equation}                
where use has been made of the expression (4.10) for the single-particle
Green's function.

The proof that use of Eq. (4.7) leads to the same result requires some
work. By taking the derivative with respect to T of Eq. (4.12) we have
\begin{equation}
{\partial \mu\over \partial T} = {1\over T} \left[ \mu +4t{V_1\over
V_0}\right]
\end{equation}        
where
\begin{equation}
V_n = {\Omega\over (2\pi)^2} \int_{\Omega_B} d^2 k {[\alpha({\rm \bf
k})]^n\over
\cosh^2 (E/2k_B T)}
\end{equation}              

By considering that the derivative with respect to $n$  of Eq. (4.12)
gives
\begin{equation}
{V_1\over V_0} =   {\int_{\Omega_B} d^2 k\alpha ({\rm \bf k}){\partial
T({\rm \bf k})\over \partial n} \over
\int_{\Omega_B} d^2 k  {\partial T({\rm \bf k})\over \partial n}} =
-{\Omega\over (2\pi)^2} {\partial\over \partial n} \int_{\Omega_B}
d^2 k\alpha ({\rm \bf k})T({\rm \bf k})
\end{equation}                      
substitution of (4.17) into (4.8) leads to
\begin{equation}
E(T,n)={4t\Omega\over  (2\pi)^2}\int^n_0 {\partial\over \partial n'}
\left[\int_{\Omega_B}
d^2 k\alpha ({\rm \bf k})T({\rm \bf k})\right] dn' =  {4t\Omega\over
(2\pi)^2}\int_{\Omega_B} d^2 k\alpha ({\rm \bf k})T({\rm \bf k})
\end{equation}                    
This concludes the proof that in the non-interacting case all
expressions (4.2), (4.3) and (4.8) give the same result. We also note
that the specific heat can be calculated by means of the following
expression
\begin{equation}
C(T,n) = -{8t^2\over k_BT^2}\left[{V^2_1\over V_0} - V_2\right]
\end{equation}        

 In Figs. 1 and 2 the linear coefficient of the electronic specific heat
$\gamma(T,n)$ is plotted as a function of temperature and filling,
respectively.

{\it a} For a fixed filling  $\gamma(T,n)$ first increases as a function
of T, exhibits a maximum at a certain temperature $T_m$, and then
decreases. The value of $T_m$ decreases by increasing the filling.  At
half-filling $T_m$ is zero and $\gamma(T,n)$ diverges as $T\to 0$; this
is an effect of the van Hove singularity (vHs). The temperature behavior
of $\gamma(T,n)$ is similar to the one exhibited by the static uniform
spin magnetic susceptibility $\chi_0$ [12,25], whereas the doping
dependence is different.

{\it b1.} At $T=0$, $\gamma(T,n)$ increases by increasing the filling,
diverges at $n=1$ and then decreases.

{\it b2.} At finite temperature the peak splits in two peaks, symmetric
with respect to $n=1$. The distance between the two peaks increases by
increasing T.
\begin{figure}[htb]   
\centerline{\psfig{figure=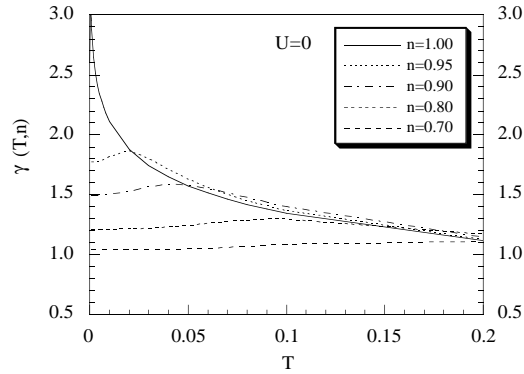,height=6.77cm,width=8.4cm}}
\caption{The linear coefficient of the normal state specific heat
$\gamma(T,n)$ of the non-interacting 2D Hubbard model is plotted as a
function of the temperature for various values of the particle density.}
\end{figure}

This is shown in Fig. 2, where $\gamma(T,n)$ is plotted as a function of
the filling at various temperatures. The shift of the two peaks with
respect to $n=1$ increases by increasing T.
\begin{figure}[htb]   
\centerline{\psfig{figure=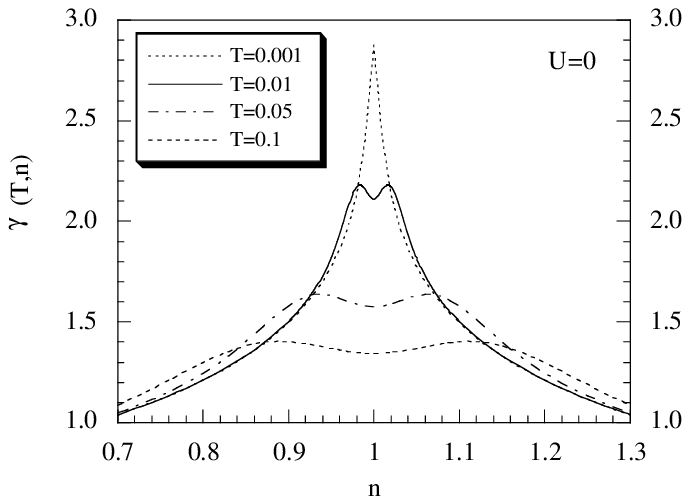,height=6.77cm,width=8.4cm}}
\caption{$\gamma(T,n)$ of the non-interacting 2D Hubbard model is
plotted as a function of the filling for various values of the
temperature.}
\end{figure}

\subsection{INTERACTING CASE}

 In the interacting case the expressions (4.2) and (4.7) for the
internal energy give different results. By recalling (2.10), Eq. (4.2)
gives
\begin{equation}
E_H=E_F - UD
\end{equation}        
where $D$  is the double occupancy. We use $E_H$ to indicate that the
internal energy has been calculated by the average of the Hamiltonian.
$E_F$ is the part of the internal energy calculated on the assumption of
only fermionic elementary excitations
\begin{eqnarray}
E_F\equiv &&\int_{-\infty}^{+\infty}d\omega N(\omega)f_F
(\omega)\omega\\
=&&{\Omega\over (2\pi)^2} \int_{\Omega_B} d^2k\left\{
\left[1-T_1({\rm \bf k})\right]\left[E_1({\rm \bf k}) +\mu\right]
A_1({\rm \bf k})+[1-T_2({\rm \bf k})][E_2({\rm \bf k}) +\mu ]A_2({\rm
\bf k})\right\}\nonumber
\end{eqnarray}             %
(4.23)

 Alternatively, we have seen that the internal energy and the specific
heat can be calculated by means of Eqs. (4.8) and (4.9). This procedure
requires a knowledge of the first and second temperature derivatives of
the chemical potential. Let us define
\begin{equation}
\mu_n = {\partial ^n\mu \over \partial T^n} \qquad \Delta_n =
{\partial^n\Delta \over \partial T^n}\qquad p_n={\partial^n p\over
\partial T^n}
\end{equation}                
By taking the first derivative with respect to T of the self-consistent
equations (2.18), we obtain
$$
G^{(1)}_0 = U(1-n) F^{(1)}_0
$$
\begin{equation}
\Delta_1={1-n\over 2}G^{(1)}_1 -{U\over 2} F_1^{(1)} +{(1-n)\over
2I_{11}I_{22}} B_1^{(1)}
\end{equation}         
$$
p_1F_1 +pF^{(1)}_1 = I_{22}F^{(1)}_1 - \Delta_1 F_0 - \Delta F^{(1)}_0
$$
where
\begin{equation}
F^{(m)}_n = {\partial ^mF_n \over \partial T^m} \qquad G^{(m)}_n =
{\partial ^mG_n \over \partial T^m}\qquad B^{(m)}_n = {\partial^mB_n
\over \partial T^m}
\end{equation}                
Explicit calculation of the derivatives defined in Eq. (4.26) shows that
the equations (4.25) provide a set of linear algebraic equations for the
three parameters $\mu_1$, $\Delta_1$, $p_1$, as functions of the
parameters  $\mu$, $\Delta$, $p$. In the same way, by taking the second
derivative with respect to T of Eqs. (2.18) we obtain
$$
G^{(2)}_0 = U(1-n) F^{(2)}_0
$$
\begin{equation}
\Delta_2={1-n\over 2}G^{(2)}_1 -{U\over 2} F_1^{(2)} +{(1-n)\over
2I_{11}I_{22}} B_1^{(2)}
\end{equation}         
$$
p_2F_1 +2p_1F^{(1)}_1 + pF^{(2)}_1 = I_{22}F^{(2)}_1 - \Delta_2 F_0 -
2\Delta_1F^{(1)}_0 - \Delta F^{(2)}_0
$$
These equations provide a set of linear algebraic equations for the
three parameters $\mu_2$, $\Delta_2$, $p_2$ as functions of the
parameters $\mu$, $\Delta$, $p$, $\mu_1$, $\Delta_1$, $p_1$. Once the
self-consistent calculation of the three parameters  $\mu$, $\Delta$,
$p$ has been performed by means of the set (2.18), then the calculation
of the first and second derivatives of the chemical potential reduces to
the solution of simple linear equations.

 In the interacting case, because of the approximation used, the
different procedures to calculate the internal energy give different
results. At first we shall compare our theoretical results with the data
obtained by numerical analysis. The specific heat $C(T)$ of the 2D
Hubbard model has been recently calculated in Ref. 34 by using quantum
Monte Carlo techniques. In particular, the Monte Carlo data for the
energy per site $E=\langle H\rangle /N$ have been fitted by polynomials
and the specific heat has been calculated by taking derivatives from
these polynomials analytically. Different polynomials have been chosen
in different regions of temperature. In the calculation of $E$ , it has
been found that finite size effects are strong at weak coupling but
become negligible for $U\ge 8$.
\begin{figure}[htb]   
\centerline{\psfig{figure=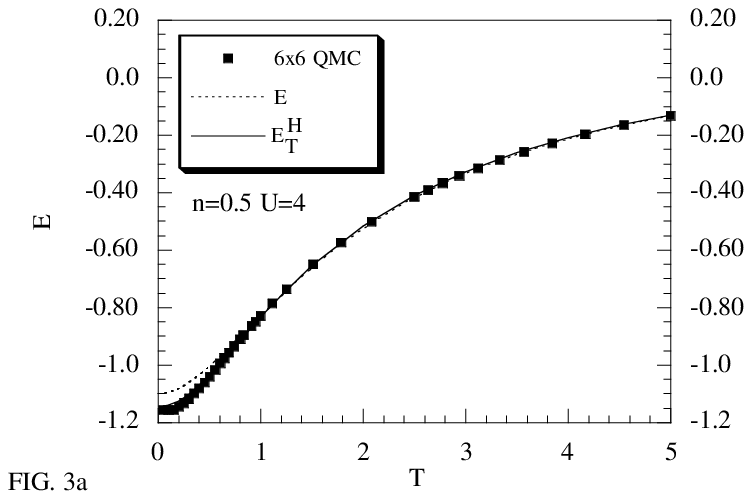,height=6.77cm,width=8.4cm}
\psfig{figure=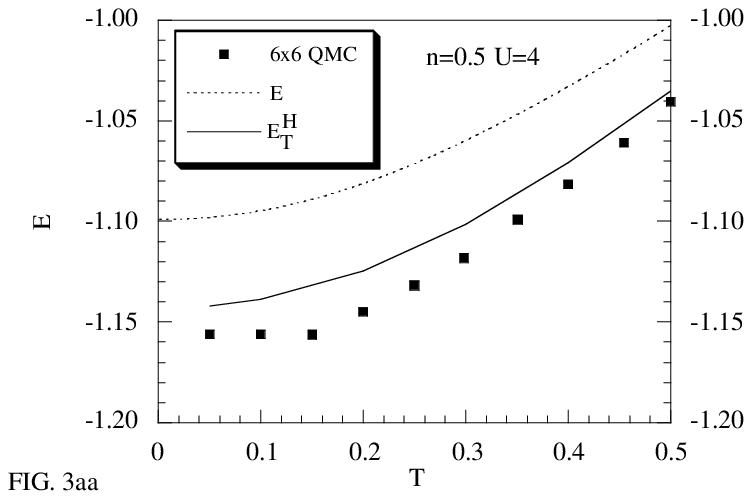,height=6.77cm,width=8.4cm}}
\centerline{\psfig{figure=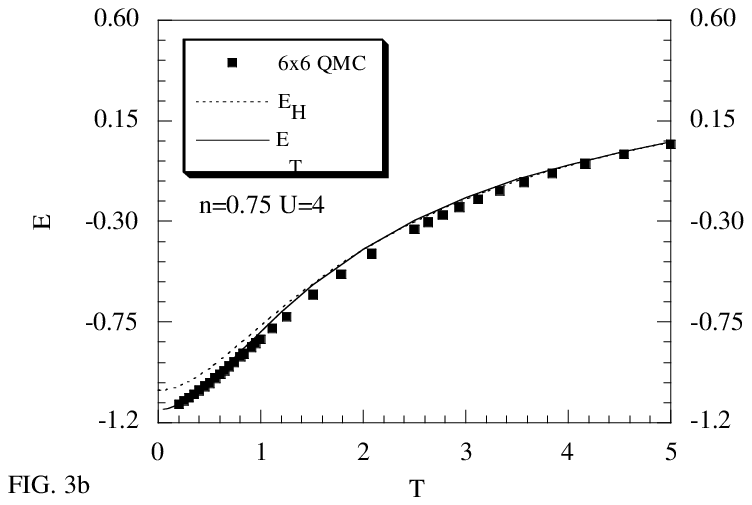,height=6.77cm,width=8.4cm}
\psfig{figure=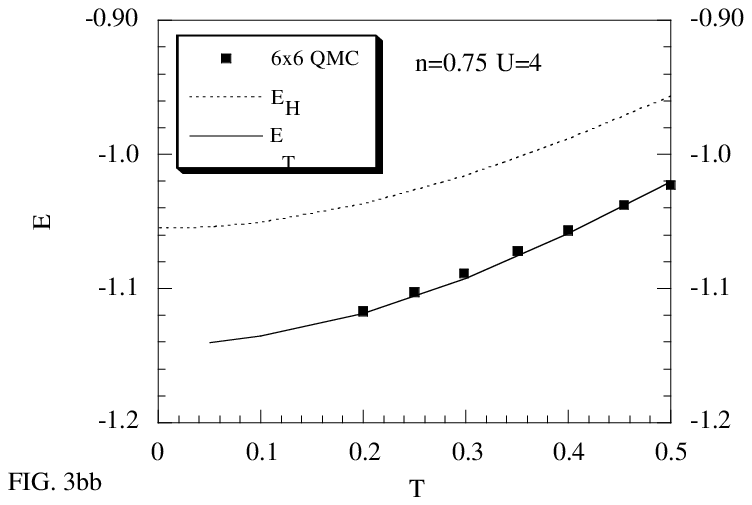,height=6.77cm,width=8.4cm}}
\centerline{\psfig{figure=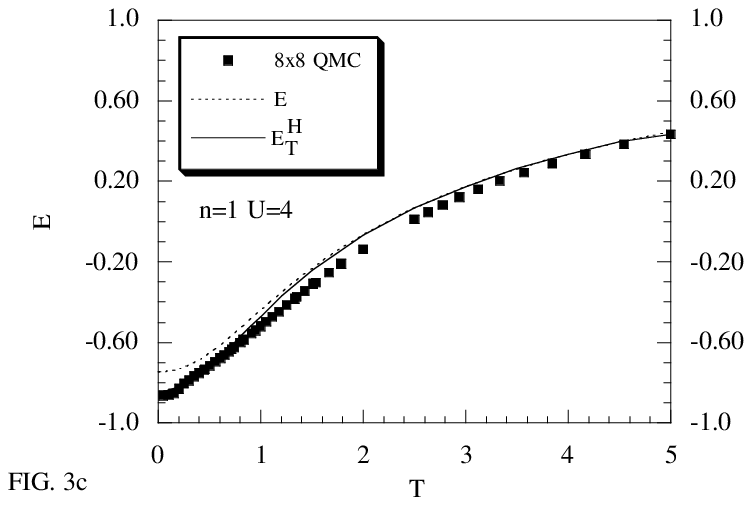,height=6.77cm,width=8.4cm}
\psfig{figure=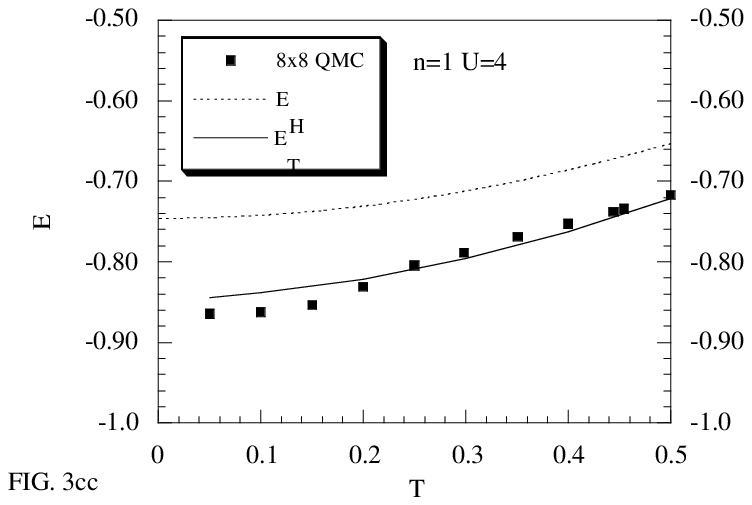,height=6.77cm,width=8.4cm}}
\caption{ The internal energy is plotted as a function of temperature
for $U=4$ and $n=0.50$ (a), $n=0.75$ (b), $n=1.0$ (c). The squares are
the QMC data of Ref. 34 for $6\times 6$ (a,b) and $8\times 8$ (c)
clusters. The dotted and solid lines refer to the theoretical results of
the COM for $E_H$, $E_T$, respectively.}
\end{figure}

In Fig. 3 we present the internal energy versus temperature in the range
$0\le T\le 5$ for several values of doping and $U=4$. The results are
compared with QMC data of Ref. 34. We are using  $E_T$
  to indicate the solution that comes from Eq. (4.8). As a general
feature we observe that $E_T$   corresponds to the lowest energy
solution and agrees very well with the QMC data, in the entire region of
temperature and for all studied dopant concentrations. When U increases,
the theoretical solution deviates from QMC in the region of low
temperatures, indicating that the antiferromagnetic (AF) correlations
are not properly taken into account in the strong coupling regime. This
is clearly seen in Fig. 4 where the internal energy is plotted versus U
for half-filling.

\begin{figure}[htb]   
\centerline{\psfig{figure=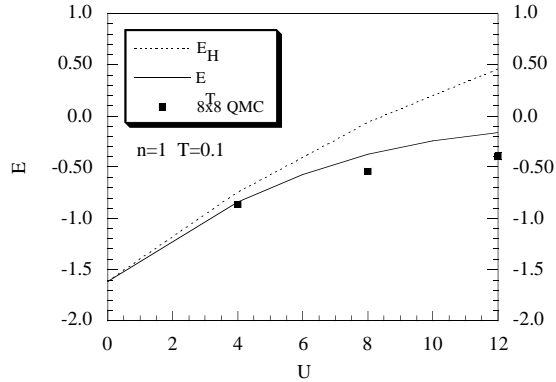,height=6.77cm,width=8.4cm}}
\caption{The internal energy is plotted as a function of $U$ for $n=1$,
$T=0.1$ (a) and $n=0.75$, $T=0.25$ (b). The squares are the QMC data of
Ref. 34 for a $8\times 8$ cluster. The dotted and solid lines refer to
the theoretical results of the COM for $E_H$, $E_T$, respectively.}
\end{figure}

We now consider the specific heat. There are two important features in
the Monte Carlo calculations: 1) a low temperature peak that appears
when the low-lying spin states are excited, and 2) a high temperature
peak which appears when states in the upper Hubbard band are excited. In
the weak coupling regime the low temperature peak moves to slightly
higher temperature as U increases, reaching a turning point at $U=7$
where the peak is at $T=0.3$. For $U>7$  the peak slowly moves to lower
temperatures, as $U$ grows. This indicates the beginning of the strong
coupling regime. The broad high temperature peak moves to higher
temperatures as $U$ increases, as expected since its presence
corresponds to the excitation of states across the gap that grows with
$U$. In the strong regime [$U>7$] the position of the charge peak
increases linearly with U: $T_{\rm charge}\approx 0.24U$. In addition,
all the curves  C  versus  T for several values of U intersect at
$T\approx 1.6$. The QMC results for the specific heat of the 2D Hubbard
model qualitatively agree with the half-filled 1D Hubbard model [53,
54].

 The specific heat $C_T=dE_T/dT$, calculated by means of Eq. (4.9), is
compared with QMC data in Fig. 5 for $U=4$ and various dopant
concentrations. At low density the agreement is generally good, in both
the weak and strong coupling cases [U=8]. At higher densities the QMC
data show a double peak structure, which is enhanced at half filling,
but also present at $n=0.75$ for $U=8$.

\begin{figure}[htb]   
\centerline{\psfig{figure=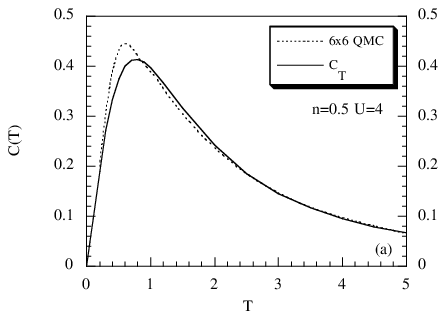,height=6.77cm,width=8.4cm}
\psfig{figure=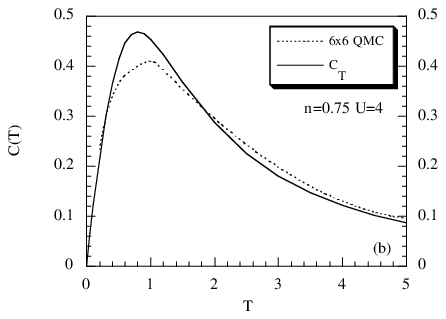,height=6.77cm,width=8.4cm}}
\psfig{figure=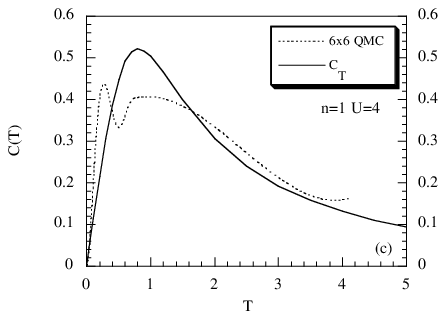,height=6.77cm,width=8.4cm}
\caption{The specific heat is plotted against temperature for $U=4$ and
$n=0.5$ (a), $n=0.75$ (b), $n=1.0$ (c). The dotted line represents the
QMC data of Ref. 34; the solid line is the result of the COM for $C_T$.
}
\end{figure}

In the discussed range of values of $U$ our results do not show a double
peak structure. The presence of two peaks in the specific heat has been
attributed to the spin and charge excitations. When U is weak the two
peaks overlap and there is no resolution. By increasing U the position
of the charge peak moves to higher temperatures and we expect to be able
to distinguish the two contributions. A study of the specific heat in
the strong coupling regime is given in Figs. 6 and 7, where the two
expressions $C_H=dE_H/dT$ and  $C_T=dE_T/dT$ are plotted, respectively,
as functions of T at half-filling.

\begin{figure}[htb]   
\centerline{\psfig{figure=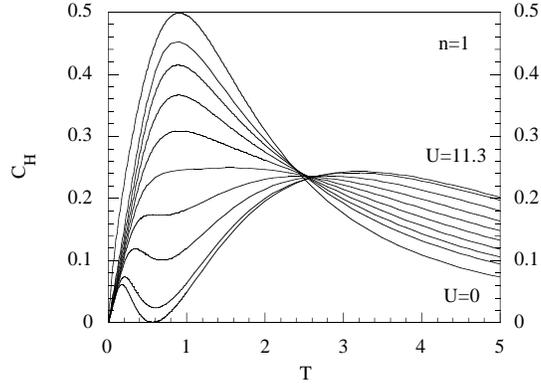,height=6.77cm,width=8.4cm}}
\caption{The specific heat $C_H=dE_H/dT$ is plotted against temperature
for half-filling and U varying in the range $0\le U\le 11.3$}
\end{figure}

\begin{figure}[htb]   
\centerline{\psfig{figure=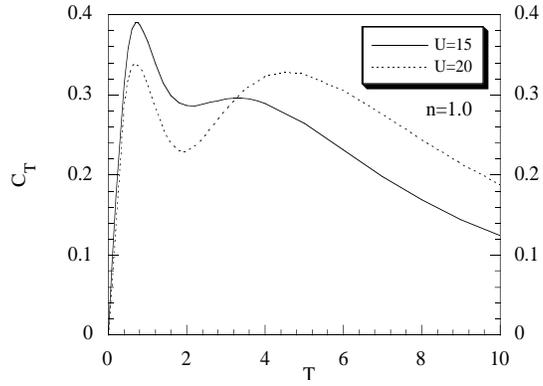,height=6.77cm,width=8.4cm}}
\caption{The specific heat  $C_T=dE_T/dT$ is plotted against temperature
for half-filling and $U=15,20$.}
\end{figure}

A peak appears at low temperatures when U is rather large [say $U\ge 8$
for $C_H$ and $U\ge 15$ for $C_T$]. This behavior qualitatively
reproduces the QMC results. In order to isolate the spin excitations for
lower values of U we can subtract from $C_T$  the contribution coming
from the pure fermionic part, $C_F=dE_F/dT$, responsible in large extent
for the charge peak structure. This is done in Fig. 8 where we report
 $C_T-C_F$ as a function of T for various values of $n$ and $U$. A
low-energy excitation appears and its contribution generally increases
by increasing $n$ and $U$.

\begin{figure}[htb]   
\centerline{\psfig{figure=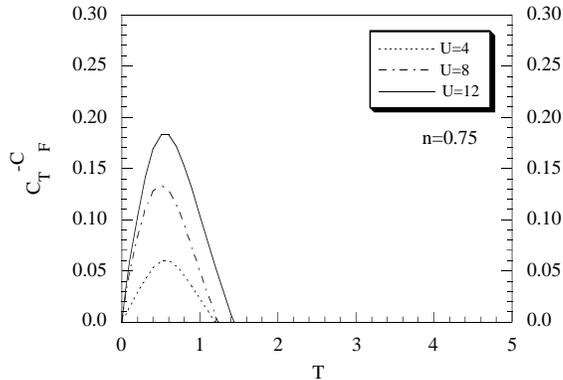,height=6.77cm,width=8.4cm}}
\caption{The contribution of the low-lying energy excitations to the
specific heat, calculated by $C_T-C_F$, is given versus T for $U=4, 8,
12$ and $n=1$.}
\end{figure}

 In Ref. 34 the position of the charge peak $T_{\rm charge}$ has been
calculated for different values of U. In our analysis when U is large
$T_{\rm charge}$ can be calculated from $C_T$ and the results lie on the
line $T_{\rm charge}=0.24U$. When U is lower, it is not possible to
resolve the two peaks. Since the charge excitation mainly comes from the
interband fermionic excitations, we can estimate $T_{\rm charge}$ from
the calculation of $C_F$. This study is shown in Fig. 9, where $T_{\rm
charge}$ is plotted versus U. The solid line is the position of the
charge peak, calculated by means of $C_F$; the triangles denote $T_{\rm
charge}$ calculated by means of $C_T$; the squares are the QMC results.

\begin{figure}[htb]   
\centerline{\psfig{figure=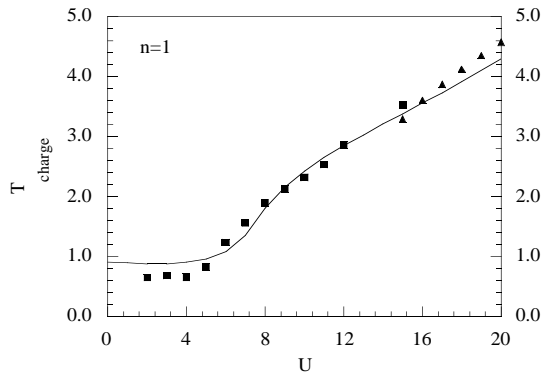,height=6.77cm,width=8.4cm}}
\caption{The position of the charge peak $T_{\rm charge}$ is plotted
against U for half-filling. The squares are the QMC data of Ref. 34 for
a $8\times 8$ cluster. The solid line and the triangles indicate $T_{\rm
charge}$ calculated by means of $C_F$ and $C_T$, respectively.}
\end{figure}

 The linear coefficient of the specific heat as a function of the
particle density has been studied by QMC for $U=8$ and T varying from
0.5 to 3. Unfortunately, due to the sign problem it is not easy to study
low temperatures in QMC. A study of $\gamma(x,T)$, by means of $C_T$,
shows a good agreement for high temperatures, but not at lower
temperatures, where QMC results exhibit a strong downward deviation in
the region of filling where we should expect an increase of
$\gamma(x,T)$ due to the effect of the van Hove singularity.

 Up to this point we have performed a detailed and ample comparison with
QMC, generally finding a quite reasonable agreement in a large region of
values for the model parameters. For intermediate values of U [U=4] the
agreement is quite satisfactory. On this basis we are confident that the
approximation used is adequate for the 2D Hubbard model and we can pass
to examine the next question as to which extent the physics of real
systems is retained in the model.

 One feature present in a large variety of systems is a characteristic
crossing point in the specific heat curves versus T [42-51]. This
behavior has been also found in 1D models [53-55] and in the Hubbard
model in infinite dimension [19], where a crossing temperature $T=0.59$
has been observed in the range $0.5\le U\le 2.5$. For the 2D Hubbard
model the same behavior, as predicted by Vollhardt [42] has been
observed by means of quantum Monte Carlo calculations [34], where for
the case of half-filling a crossing temperature $T=1.6\pm 0.2$ has been
observed in the range $2\le U\le 12$. In Fig. 10 the specific heat $C_T$
is given versus T for half-filling and various values of U. When $n=1$
the curves cross at the same temperature $T\approx 2.0$; when doping is
considered the region of crossing spreads out and moves to higher
temperatures. From the thermodynamic relations, this crossing
temperature $T_U$ corresponds to a turning point of the double occupancy
as a function of $T$, that is $(\partial^2D/\partial T^2)_{T_U}=0$. A
study of the function $(\partial^2D/\partial T^2)$  will be done in the
next Section.

\begin{figure}[htb]   
\centerline{\psfig{figure=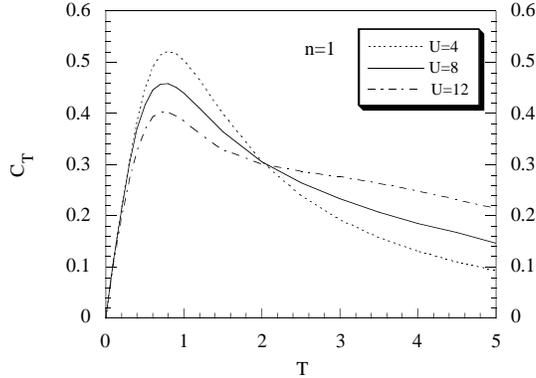,height=6.77cm,width=8.4cm}}
\caption{The specific heat $C_T= dE_T/dT$ is plotted against temperature
for different values of U and $n=1$.}
\end{figure}

 As mentioned in Section III, the linear coefficient of the specific
heat of cuprates exhibits an anomalous behavior in the normal state. To
investigate this, in Fig. 11 we present the linear coefficient
$\gamma(x,T)$ as a function of the doping $x=1-n$  for various
temperatures. As a general behavior we see that by increasing the doping
$\gamma(x,T)$ increases up to a certain doping and then decreases. The
nature of the peak is due to the fact that the Fermi energy crosses the
vHs for a certain critical value $x_c$. The value of $x_c$ depends on
the ratio U/t and varies between 0 and 1/3, as U increases from zero to
infinite. For $U/t =4$ it is found $x_c=0.27$, very close to the
experimental value observed in $La_{2-x}Sr_x CuO_4$  [37, 38]. At
half-filling the Fermi energy ($\epsilon_F$) is at the center of the two
Hubbard bands; by varying the dopant concentration some weight is
transferred from the upper to the lower band, $\epsilon_F$  moves to
lower energies and crosses the vHs for a critical value of the doping;
increasing $x$, further moves $\epsilon_F$ away from the vHs. A study of
the Fermi surface shows that for $x>x_c$ we have a closed surface which
becomes nested at  $x=x_c$  and opens for $x<x_c$. An enlarged Fermi
surface with a volume larger than the non-interacting one has been
reported by QMC calculations [56, 57] and by other theoretical works
[58, 59].

\begin{figure}[htb]   
\centerline{\psfig{figure=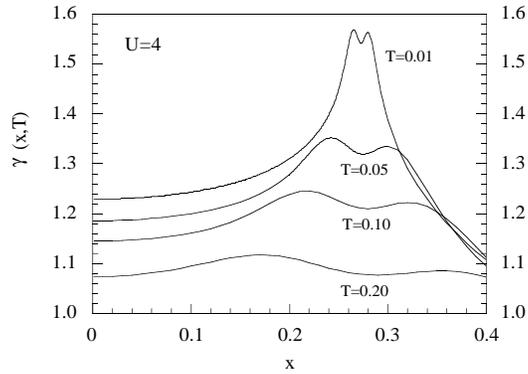,height=6.77cm,width=8.4cm}}
\caption{The linear coefficient of the specific heat $\gamma(x,T)$,
calculated from $C_T$, is given as a function of the doping  $x=1-n$ for
$U=4$ and different temperatures.}
\end{figure}

\begin{figure}[htb]   
\centerline{\psfig{figure=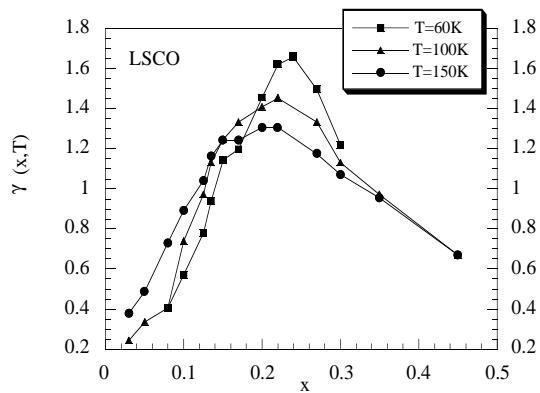,height=6.77cm,width=8.4cm}}
\caption{The linear coefficient of the specific heat $\gamma(x,T)$ for
$La_{2-x}Sr_x CuO_4$ is shown as a function of the $Sr$ content. The
dots are the experimental data for different temperatures, taken from
Refs. 37 and 38.}
\end{figure}

The peak position of $\gamma(x,T)$ depends on the temperature. In the
limit of zero temperature a sharp peak is exactly located at $x=x_c$. By
increasing the temperature the peak moves away from $x_c$ and broadens
into two peaks. The situation is similar to what we have calculated for
the non-interacting case [cfr. Fig. 2]; we find that the role played by
the interaction manifests itself through the shift of the vHs and the
band structure which creates an asymmetry between the two peaks. The
behavior described in Fig. 11 well reproduces the experimental
situation.

In the case of $La_{2-x}Sr_x CuO_4$ the experimental data [37, 38] for
$\gamma(x,T)$ are reported in Fig. 12. The peak exhibited by
$\gamma(x,T)$ decreases in intensity and moves to lower values of doping
when T increases. In the case of $YBa_2Cu_3O_{6+y}$, the experimental
results reported in Ref. 39 are for the higher temperature $T= 280K$;
$\gamma(x,T)$ increases with doping and presents two broad maxima in the
region of high doping.

 Interpretation of the experimental results obtained in Ref. 39 for
$YBa_2Cu_3O_{6+y}$ in terms of a sharp feature in the density of states,
consistent with ARPES experiments [60], was presented in Refs. 61 and
62. The consistency of thermodynamic data with the presence of a vHs
near the Fermi level was shown in Ref. 58 by considering a p-d
like-model in the framework of slave-boson mean-field theory in the
limit of large U.

 We also mention that the specific heat of $La_{2-x}Sr_x CuO_4$ has been
estimated in Ref. 40 from the data for the heat capacity anomaly at the
superconducting transition temperature by assuming a BCS-type relation.
Under this assumption the authors find the same behavior for
$\gamma(x,T)$, but with a peak located around $x\approx 0.18$, close to
the optimal doping.

\begin{figure}[htb]   
\centerline{\psfig{figure=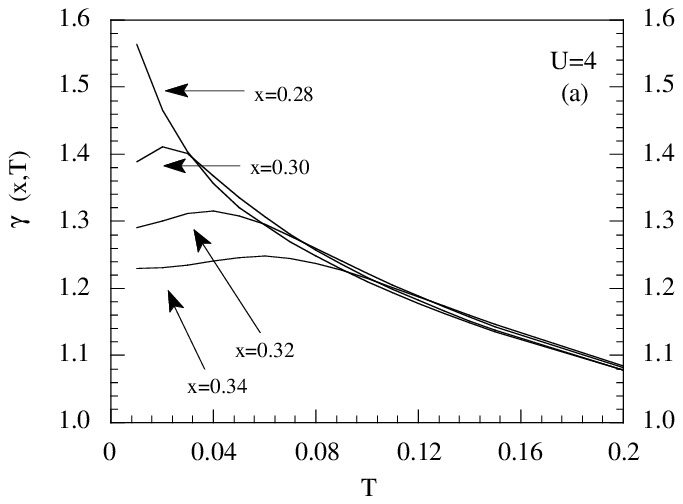,height=6.77cm,width=8.4cm}
\psfig{figure=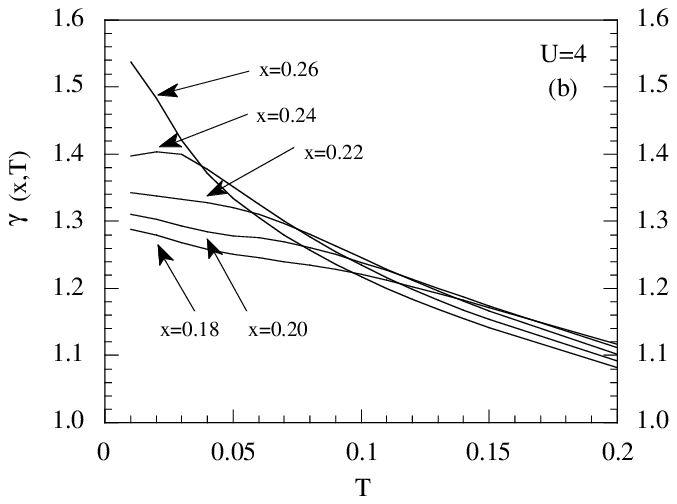,height=6.77cm,width=8.4cm}
}
\caption{The linear coefficient of the specific heat  $\gamma(x,T)$,
calculated from $C_T$, is given as a function of temperature for U=4. In
Figs. 19a and 19b the curves have been traced for $x>x_c$ and $x<x_c$,
respectively.}
\end{figure}

 In Figs. 13a and 13b we present the linear coefficient $\gamma(x,T)$ as
a function of the temperature for values of the filling $x>x_c$ and
$x<x_c$, respectively. At $x=x_c$ we see that $\gamma(x,T)$ diverges as
$T\to 0$; this is an effect of the vHs. When $x\neq x_c$ the Fermi
energy moves away from the vHs and the peak exhibited by $\gamma(x,T)$
moves away from T=0. As shown in Figs. 13a and 13b, $\gamma(x,T)$ as a
function of temperature has different behaviors in the two regions
$x>x_c$ and $x<x_c$. In the overdoped region $\gamma(x,T)$ firstly
increases as a function of T, exhibits a maximum at a certain
temperature $T_m$ and then decreases. This behavior is similar to the
one exhibited by $\chi_0(T)$ [12,25]. As shown in Fig. 13a, when the
doping decreases the value of $T_m$ moves to lower temperatures. This
behavior qualitatively agrees with the non interacting case showing that
for $x>x_c$ the AF correlations are weak. A different situation is
observed in the underdoped region, where $\gamma(x,T)$ is always a
decreasing function of T. When we look at the experimental results for
$La_{2-x}Sr_x CuO_4$ [37, 38] and for $YBa_2Cu_3O_{6+y}$ [39, 52] we
find that the behavior of $\gamma(x,T)$ as a function of T in the
underdoped region is more similar to that for the non-interacting case,
showing that cuprate materials could be conveniently described in terms
of a Fermi liquid picture, dominated by the Van Hove singularity. Then,
if we want to use the Hubbard model to describe high-$T_c$ cuprates, we
might assume a scheme for non-interacting -like particles and use Eq.
(4.23) to calculate the specific heat. When we do this, we find the
results reported in Figs. 14, 15a and 15b.

\begin{figure}[htb]   
\centerline{\psfig{figure=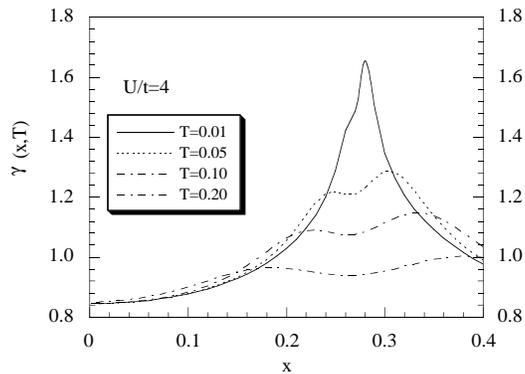,height=6.77cm,width=8.4cm}}
\caption{The linear coefficient of the specific heat $\gamma(x,T)$,
calculated from $C_F$, is given as a function of the doping $x=1-n$ for
$U=4$ and different temperatures.}
\end{figure}

\begin{figure}[htb]   
\centerline{\psfig{figure=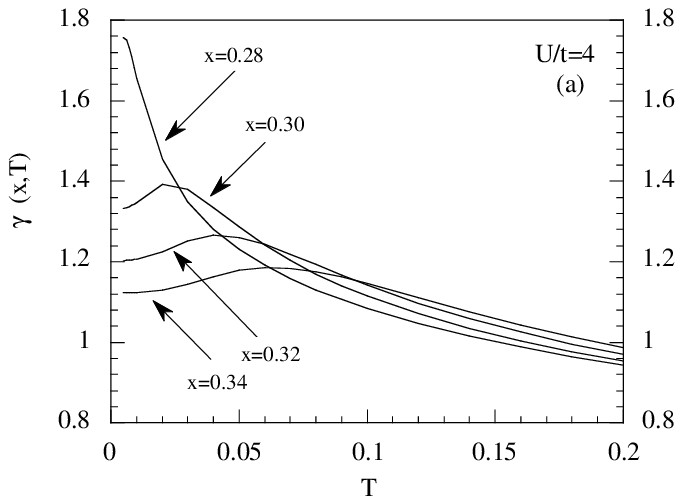,height=6.77cm,width=8.4cm}
\psfig{figure=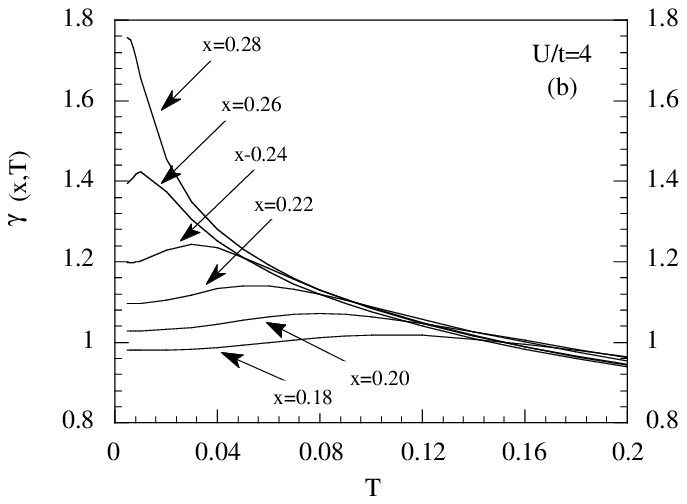,height=6.77cm,width=8.4cm}}
\caption{The linear coefficient of the specific heat $\gamma(x,T)$,
calculated from $C_F$, is given as a function of temperature for U=4. In
Figs. 15a and 15b the curves have been traced for $x>x_c$ and $x<x_c$,
respectively.}
\end{figure}

As a function of the doping $\gamma(x,T)$ has a similar behavior for
that part dominated by the Van Hove singularity, but the band structure
plays a different role. By comparing Figs. 11 and 14 it is worthwhile to
stress that the presence of a maximum in $\gamma(x,T)$ does not
necessarily indicate a pure fermionic excitation spectrum; it is a
consequence of the enhancement in the density of states when the Fermi
energy crosses the vHs.  This becomes more evident in Fig. 15a and 15b
where the temperature dependence is studied at fixed doping. Both in the
underdoped and overdoped regions $\gamma(x,T)$ exhibits a maximum at a
certain temperature. The behavior given in these figures qualitatively
reproduces the experimental situation reported in Ref. 37, 38, 39, 52.
The fact that for $YBa_2Cu_3O_{6+y}$ $\gamma(x,T)$  is always a
decreasing function of T when $y>0.8$ is understood because by
approaching the critical doping, $T_m$ is shifted to low temperatures,
below the critical superconducting temperature.

\section{THE DOUBLE OCCUPANCY}

 As a simple thermodynamic quantity indicating the degree of correlation
of the system, in this Section we study the double occupancy D, defined
as the fraction of doubly occupied sites
\begin{equation}
D=\langle n_\uparrow n_\downarrow\rangle
\end{equation}       
This quantity can be calculated by means of the expression
\begin{equation}
D={n\over 4}\left[ 1-G_0-UF_0\right]
\end{equation}      
Then, the first and second temperature derivatives of D can be
analytically calculated as
\begin{equation}
{dD\over dT}=-{n\over 4}\left[G^{(1)}_0-UF^{(1)}_0\right]\qquad
{d^2D\over dT^2}=-{n\over 4}\left[G^{(2)}_0-UF^{(2)}_0\right]
\end{equation}                
once the self-consistent equations have been solved.

 We display in Figs. 16a and 16b the temperature dependence of D for
various values of U at particle concentrations $n=0.7$ and $n=0.8$. In
Fig. 16a the data display a characteristic low temperature behavior: D
initially decreases with temperature, reaches a minimum, and increases
again. In other words, the curve indicates the presence of a T region
where the formation of local magnetic moments is enhanced with
increasing T [the double occupancy D determines the local spin-spin
correlation function $S^2$ through the equation $S^2=3(n-2D)/4$]. This
behavior is characteristic of incipient localization effects in a
strongly correlated Fermi liquid in a regime dominated by spin
fluctuations. Starting from the low temperature Fermi liquid regime,
when the temperature increases, the system can gain free energy by
localizing the particles (i.e. decreasing D) in order to take advantage
of a larger spin entropy [19,53]. In the absence of spin excitations one
would observe decreasing values with increasing T.

\begin{figure}[htb]   
\centerline{\psfig{figure=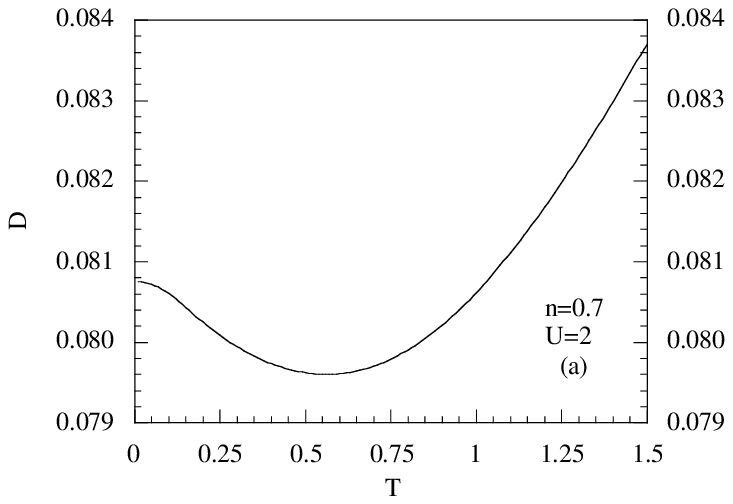,height=6.77cm,width=8.4cm}
\psfig{figure=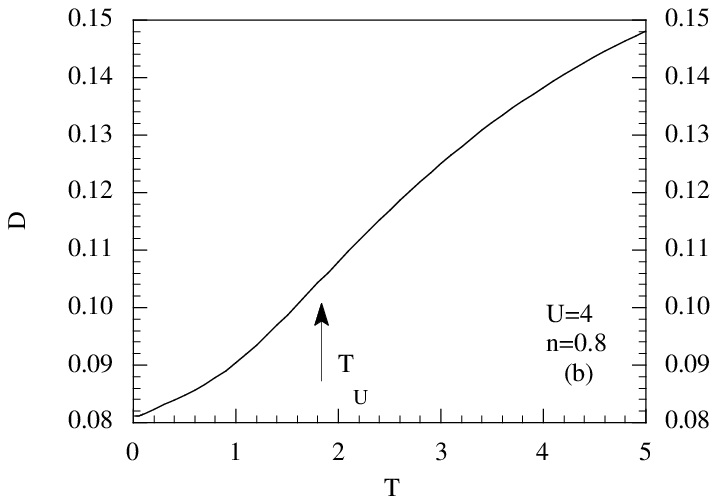,height=6.77cm,width=8.4cm}}
\caption{The double occupancy $D$  is plotted as a function of the
temperature for $n=0.7$, $U=2$ (a) and for $n=0.8$, $U=4$ (b). The arrow
in Fig. 22b indicates the temperature $T_U$ where the curve changes
curvature.}
\end{figure}

 In Fig. 16b D is a monotonic increasing function of temperature. In
this case the values of n and U are large enough to inhibit localization
effects due to the increase of temperature. To study this behavior in
more detail this behavior, the derivative with respect to the
temperature of the double occupancy has been analyzed. The results show
that for a given T there exists a critical value of U, say $U_D(T)$,
such that
\begin{equation}
{\partial D\over \partial T}<0 \qquad {\rm for}\qquad U<U_D(T)
\end{equation}              
\begin{equation}
{\partial D\over \partial T}>0 \qquad {\rm for}\qquad U>U_D(T)
\end{equation}              

The function  $U_D(T)$, defined by $\partial D/\partial T=0$, is given
in Fig. 17. We note that at T=0  $U_D(0)$ coincides with $U_c(n,0)$,
defined as the critical strength of the on-site Coulomb interaction for
which the Fermi energy crosses the vHs at fixed dopant concentration.
$U_D(T)$ goes to zero for some temperature $T_{D}$. For $n=0.7$ we find
that $T_D=0.581$. When  $T>T_D$ we have $\partial D/\partial T>0$ for
all values of U. The behavior of $T_D$  as a function of $n$ is reported
in the Fig. 18. The fact that $(\partial D/\partial T)_{U=U_D(T)}=0$
implies that at $U=U_D(T)$ the double occupancy does not depend on T.
However, the curves of D  as function of U for different values of T
will not cross in a single point because $U_D(T)$ changes with the
temperature in a significant way.

\begin{figure}[htb]   
\centerline{\psfig{figure=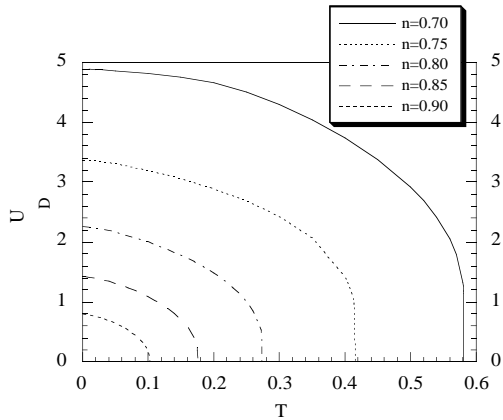,height=6.77cm,width=8.4cm}}
\caption{ $U_D(T)$ as a function of the temperature for various values
of the filling.}
\end{figure}

 At very high temperature $T\gg U$, larger than $T_U$ where there is a
change in the concavity, D asymptotically tends to the atomic value
$n^2/4$, as expected. We have seen in the previous section that the
specific heat curves versus T for different values of U cross almost at
the same point $T_U$, determined by the equation
\begin{equation}
\left({\partial ^2D\over \partial T^2}\right)_{T_U}=0
\end{equation}    
 \begin{figure}[htb]   
\centerline{\psfig{figure=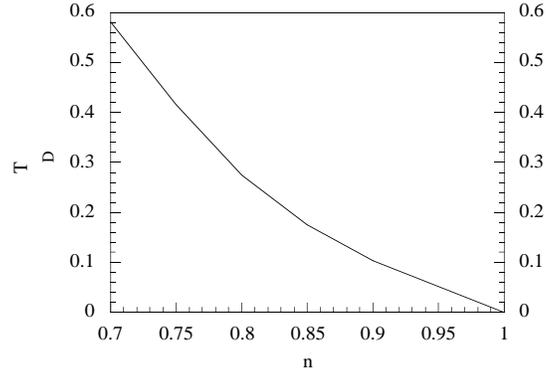,height=6.77cm,width=8.4cm}}
\caption{ $T_D$ as a function of the filling.}
\end{figure}

A study of this equation by means of formula (5.3) gives the results
plotted in Fig. 19, where $T_U$ is plotted versus U for $n=0.75$.
 \begin{figure}[htb]   
\centerline{\psfig{figure=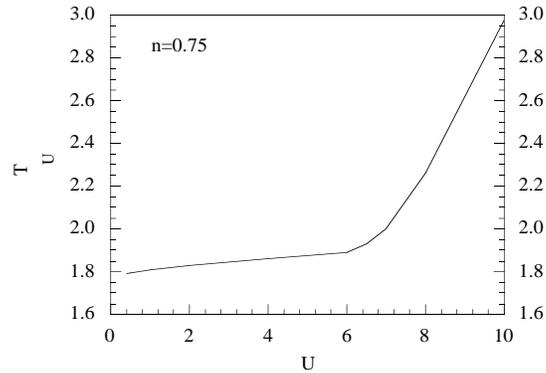,height=6.77cm,width=8.4cm}}
\caption{$T_U$ as a function of the potential intensity U for $n=0.75$.}

\end{figure}

\section{CHEMICAL POTENTIAL VERSUS TEMPERATURE}

 From the solution of the fermionic propagator and by means of Eqs.
(4.25) we obtain the following behavior for the temperature derivative
of the chemical potential
\begin{eqnarray}
{\partial\mu\over \partial T}&<& 0\qquad {\rm for}\qquad
n<n_\mu(T)\nonumber\\ {\partial\mu\over \partial T}&=& 0\qquad {\rm
for}\qquad n=n_\mu(T)\\ {\partial\mu\over \partial T}&>& 0\qquad {\rm
for}\qquad n>n_\mu(T)\nonumber
\end{eqnarray}     

\begin{figure}[htb]   
\centerline{\psfig{figure=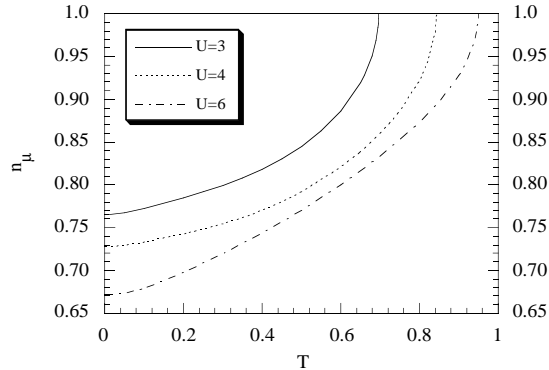,height=6.77cm,width=8.4cm}}
\caption{$n_\mu(T)$ as a function of the temperature for different
values of U.}
\end{figure}
The function $n_\mu(T)$ is presented in Fig. 20. It can be shown that at
$T=0$ $n_\mu$ coincides with $n_c$, the critical value where the Fermi
level crosses the van Hove singularity. But, the temperature dependence
of $n_\mu(T)$ is remarkably different from the one of  $n_c(T)$.
\begin{figure}[htb]   
\centerline{\psfig{figure=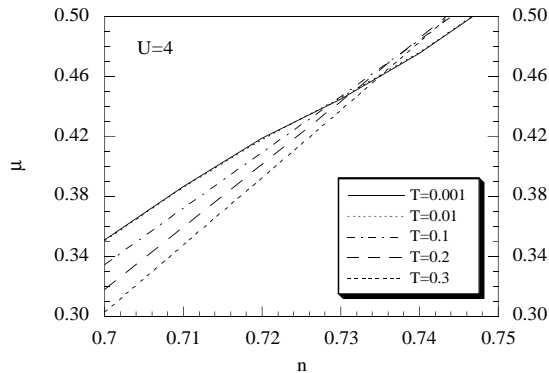,height=6.77cm,width=8.4cm}}
\caption{The chemical potential $\mu$  as a function of the filling for
$U=4$ and different temperatures.}
\end{figure}
Thus, only for $T\to 0$ we may relate the transition
$\partial\mu/\partial T<0
\Longrightarrow \partial\mu\partial T>0$ to the reversal of the
sign of the derivative for the density of states at the Fermi level.
Furthermore, we observe that  $n_\mu(T)$ reaches the value of 1  for
some temperature $T_\mu$   (for $U=4$ we find $T_\mu=0.843$). When
$T>T_\mu$ we have $\partial\mu/\partial T<0$ for all values of $n$ . The
fact that $(\partial\mu/\partial T)_{n=n_\mu}(T)=0$ implies that at
$n=n_\mu(T)$  the chemical potential does not depend on T. Therefore,
the curves of $\mu$ as function of $n$ for various values of $T$,
reported in Fig. 29, will not cross exactly in the same point as claimed
in Ref. 63.

\section{THE ENTROPY}

  The entropy $S(T,n)$, connected to the total number of spin and charge
excitations at temperature T and filling $n$, is a bulk thermodynamic
quantity uniquely determined by the spectrum of excitations, whose
magnitude and temperature dependence provide an important test for
proposed theories. Theoretical works available so far are the following.
Bipolaron models propose preformed boson charge carriers at $T_c$ and
behaving classically at higher temperatures. Apart from some
inconsistency related to the magnitude of the entropy, these theories
have to resort to the existence of thermally excited triplet bipolarons
in explaining the deviation of S from a linear T-dependence in
underdoped samples [64]. Theoretical studies of the entropy in the
strongly correlated electron systems have also been performed in the
framework of the statistical spin liquid [65-68]. This scheme is based
on the assumption that in the strongly correlated metals the doubly
occupied single-spin configurations must be excluded not only in the
real space representation, but also in the reciprocal space. By means of
the spin liquid statistics, the entropy of localized moments is
reproduced when the Mott insulator limit is reached for half-filling.
Nevertheless, this over imposed statistics freezes the system in a wrong
Hilbert space whenever different choices of the parameters modify the
interplay between thermal excitations and electronic interactions. Some
theories [3] predict decoupled holon (boson) and spinon (fermion)
excitations. In these approaches it is difficult to reconcile the
experimentally observed magnitude for the entropy with its partition
between statistically independent excitations. Moreover, the striking
numerical correlation between $S/T$ and $a\chi_0$ is expected for weakly
interacting fermions but not if the dominant excitations are those of
spinless bosons. In Ref. 63 exact diagonalization studies of the $t-J$
model have been performed; for several thermodynamic quantities a
critical doping concentration that marks a change of the Fermi surface
character is found.

 By means of the relation (4.7), we have calculated the entropy per site
$S(T,n)$. Recalling the behavior of $\partial\mu/\partial T$, we see
that the entropy must have the following behavior

\noindent (i) for  $T<T_\mu$, $S(T,n)$  increases with increasing
particle concentration, reaches a maximum for $n=n_\mu$, then decreases
(see Fig. 22a);

\noindent (ii) for $T>T_\mu$, $S(T,n)$  always increases with increasing
particle concentration (for $U=4$ we find $T_\mu=0.843$) as is shown in
Fig. 22b.

\begin{figure}[htb]   
\centerline{\psfig{figure=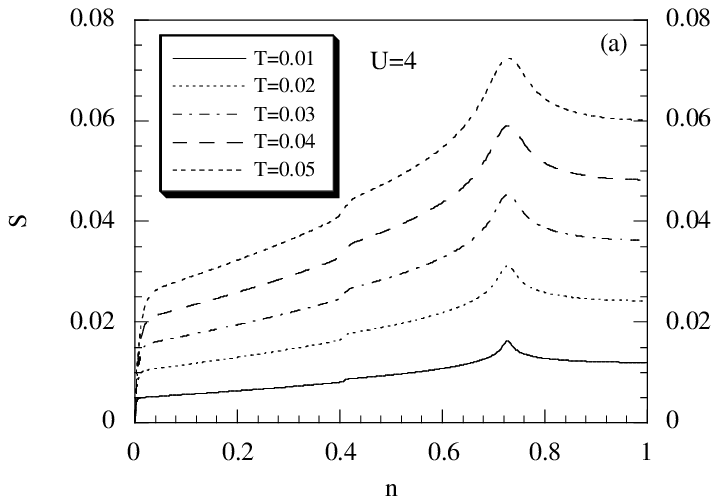,height=6.77cm,width=8.4cm}
\psfig{figure=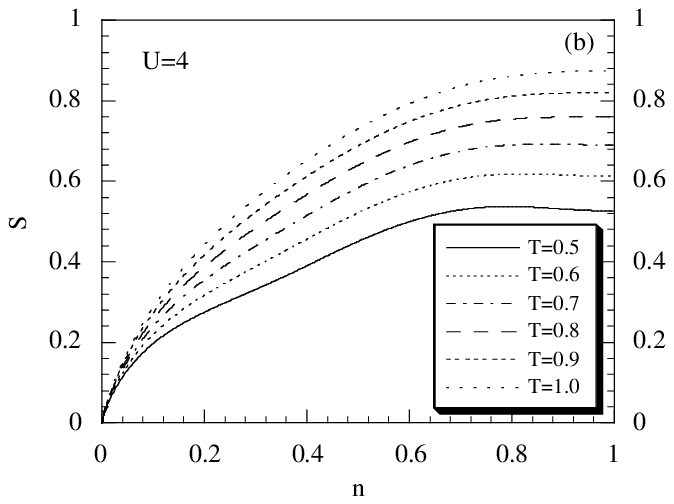,height=6.77cm,width=8.4cm}}
\caption{The entropy is plotted versus the particle density for $U=4$
and different temperatures.}
\end{figure}

Again the peak structure reflects a Fermi level crossing the vHs at the
critical doping. This behavior is in agreement with the experimental
data from Refs. 38, 39 and 52. Indeed, the experiments show a well
defined peak structure in a large region of temperature (from 40K to
320K); furthermore, the position of the peak slightly changes with
temperature. In the theoretical analysis the position of the peak as a
function of temperature is governed by $n_\mu(T)$, reported in Fig. 20,
that shows a smooth variation in the region of physical relevance
($T<0.05$). In Fig. 23 the entropy versus the particle concentration is
reported for various values of the Coulomb interaction.

\begin{figure}[htb]   
\centerline{\psfig{figure=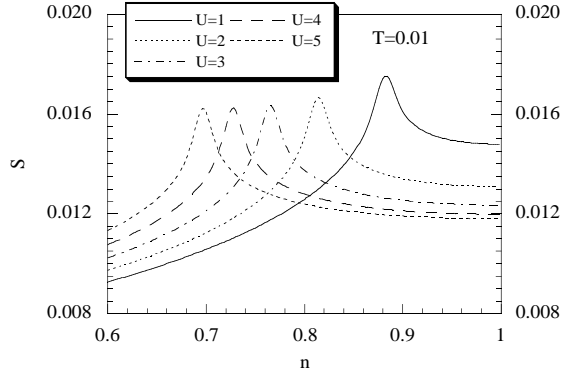,height=6.77cm,width=8.4cm}}
\caption{The entropy is plotted versus the particle density for $T=0.01$
and different values of U.}
\end{figure}

The maximum of the entropy shifts to lower values of $n$ by increasing
U, varying between 1 and 2/3 when U varies from zero to $\infty$. In
Fig. 24, we see that for $T=0.4$ all entropy curves for different U
cross at the same particle concentration. By comparison with figure 21
and by numerical analysis, we see that the crossing point is exactly the
critical concentration $n_\mu(T)$. Recalling the Maxwell relation
\begin{equation}
\left({\partial S\over \partial U}\right)_T = -\left({\partial D\over
\partial T}\right)_U
\end{equation}       
the behavior shown in Fig. 22 implies $(\partial D/\partial
T)_{n=n_\mu}=0$. If we remember that the double occupancy $D$ determines
the local spin-spin correlation function, it is clear that a sign
reversal of its derivative with respect to the temperature represents a
crossover from a regime dominated by spin fluctuations, where S is a
decreasing function of U, to another regime favouring charge
fluctuations (electronic delocalization), where the entropy is an
increasing function of U.
 \begin{figure}[htb]   
\centerline{\psfig{figure=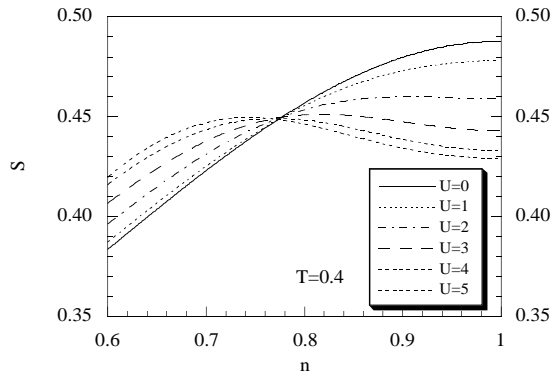,height=6.77cm,width=8.4cm}}
\caption{The entropy is reported versus the particle density for $T=0.4$
and different values of U.}
\end{figure}

In Figs. 25 and 26 we report the temperature dependence of $S$ and $S/T$
for several dopant concentrations. The curves have a qualitative
agreement with the experimental ones [38,39,52]. In Fig. 25a, where
$x>x_c$, S is a decreasing function of $x$ at a fixed temperature; the
opposite behavior is observed in Fig. 25b.
 \begin{figure}[htb]   
\centerline{\psfig{figure=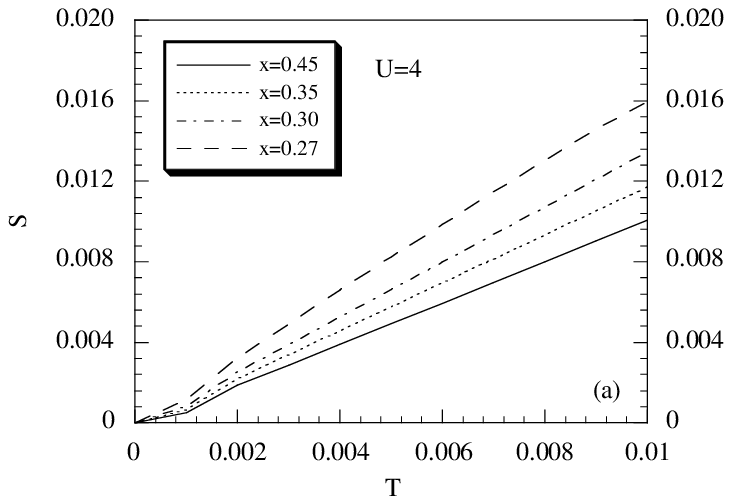,height=6.77cm,width=8.4cm}
\psfig{figure=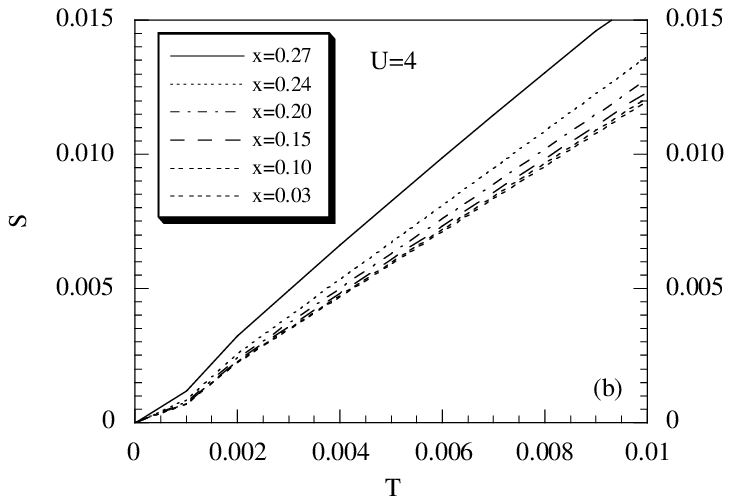,height=6.77cm,width=8.4cm}}
\caption{The entropy is plotted versus the temperature for $U=4$. The
range of filling is $x>x_c$ for (a) and $x<x_c$ for (b)}
\end{figure}

In the limit of zero temperature the entropy goes to zero by a linear
law. When T increases the entropy deviates from the linear behavior. In
the region $0.01\le T\le 0.1$ the temperature dependence is well
described by the law
\begin{equation}
S(T)=S_0+S_1T+S_2T^2
\end{equation}           
where the coefficients $S_0$, $S_1$, $S_2$ are strongly dependent on the
filling.
 \begin{figure}[htb]   
\centerline{\psfig{figure=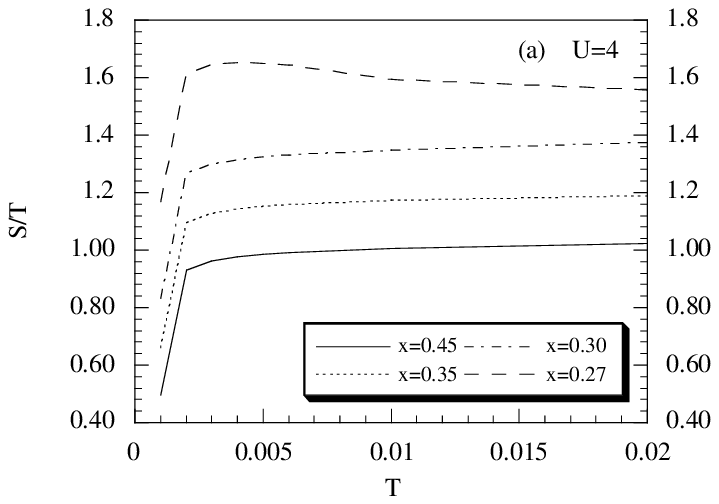,height=6.77cm,width=8.4cm}
\psfig{figure=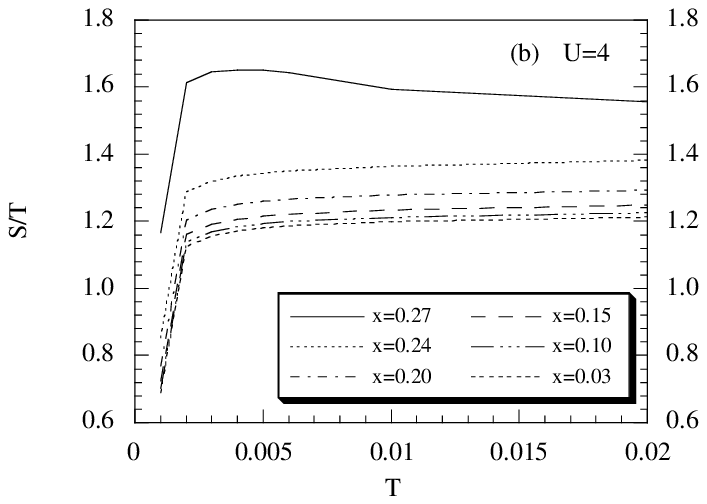,height=6.77cm,width=8.4cm}}
\caption{ $S/T$   is plotted versus the temperature for $U=4$. The range
of filling is $x>x_c$ for (a) and $x<x_c$  for (b)}
\end{figure}

 It is worth noticing that in the limit of large temperatures (a weak
point of the statistical spin liquid [65-68]) our results for the
entropy asymptotically agree with the exact expression
\begin{equation}
\lim_{T\to\infty} S(T,n)=2\ln 2-n\ln n - (2-n)\ln(2-n)
\end{equation}                

 For non-interacting fermions at $T=0K$ we have
\begin{equation}
\gamma=a\chi_0
\end{equation}    
where $\chi_0$ is the bulk susceptibility and $a$ is the Wilson ratio
\begin{equation}
a={\pi^2\over 3}
\end{equation}  
In the case of $La_{2-x}Sr_x CuO_4$ [38] and of $YBa_2Cu_3O_{6+y}$
[39,52,69], there is a striking numerical correlation between $S/T$ and
$a\chi_0$. As noticed by Loram et al., this resemblance shows that the
total spin+charge spectrum over all moments ${\rm \bf k}$ (from S) and
the ${\rm \bf k}=0$ spin spectrum (from $\chi_0$) have a similar energy
dependence. Also, experimental evidences suggest that the low-energy
excitations are predominantly those of conventional fermions, and that
the substantial T dependencies of $S/T$ and $\chi_0$ are primarily
determined by the energy dependence of the single-particle density of
states in the vicinity of the Fermi level.
\begin{figure}[h]   
\centerline{\psfig{figure=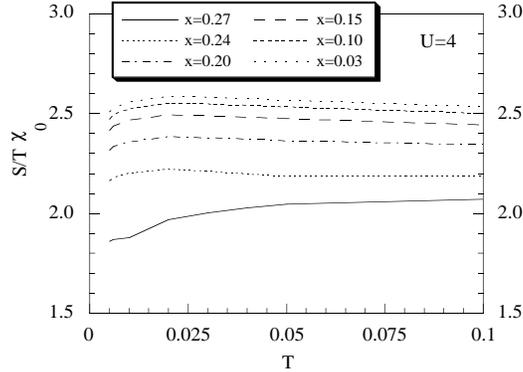,height=6.77cm,width=8.4cm}}
\caption{ $S/T\chi_0$ is plotted versus the temperature for $U=4$ and
different fillings.}
\end{figure}

 In Fig. 27 we present $S/T\chi_0$ as a function of the temperature for
various values of doping (i.e., $0.03\le x\le 0.27$). For all values of
dopant concentration $S/T\chi_0$ is almost constant over a wide range of
temperatures.

In addition, the value of $S/T\chi_0$ is less than the non-interacting
one. This is due to the fact that by introducing interaction the number
of microscopic states accessible to the same macroscopic state is
reduced (i.e., the entropy per site) whereas the susceptibility is
increased by incipient localization effects.

 In order to obtain a better understanding of how the thermodynamics of
an electronic liquid is modified by the interaction, we have performed a
study of non-interacting Hubbard model (i.e. $U=0$). What we learn from
the study of this model is that the critical value (i.e. half-filling)
above which the entropy looks a decreasing function of the filling is
uniquely fixed by the statistics. The temperature has no role. On the
contrary, in the interacting case the energy scale of charge
configurations has a crucial role in the region of particle
concentration between $n_\mu$ and half-filling. For $n_\mu<n<1$ there is
a critical temperature $T_\mu$, depending on the filling, above which
the behavior is similar to the non-interacting case [i.e. where
$\partial S/\partial n$ becomes positive, or where $\partial\mu/\partial
T$  becomes negative].

 We now consider the physical origin of these results. In the
non-interacting case the combinatorics dictated by the Fermi statistics
governs the behavior of the entropy. This can be understood if we think
that for the single-site problem the entropy has the values 0, $\ln 2$
and 0 for occupancy 0, 1  and 2, respectively. In the interacting case
it is natural to look for a critical value of the filling above which
the number of permutations satisfying the restrictions of the boundary
conditions starts to decrease. For $n_\mu<n<1$, because of the Coulomb
interaction, by increasing the particle density the number of
microscopic realizations, accessible to the same observable macroscopic
state, decreases [the Pauli principle is obviously crucial for a correct
counting]. This is true only if the thermal excitations do not exceed
the energy scale fixed by the interaction. Definitely, for $0<n<n_\mu$
we have a sort of disordered non-interacting state with $\partial
S/\partial n>0$, whereas for  $n_\mu<n<1$ the low-lying excitations
characterize a far from random spatial pattern [i.e. $\partial
S/\partial n<0$]. In the range $n_\mu<n<1$ incommensurate magnetism and
superconductivity are experimentally observed [12, 70].

\section{CONCLUDING REMARKS}

 The 2D single-band Hubbard model has been studied by means of the
composite operator method. By considering the Hubbard operators as basic
set of fields, which describe interatomic excitations restricted to
subsets of the occupancy number, the single-particle electronic
propagator has been computed in a fully self-consistent way by means of
a quasi-particle scheme capable of coherently integrating dynamics,
boundary conditions and symmetry principles.

 The paper was devoted to the study of the electronic specific heat and
entropy per site in the paramagnetic phase. We analyzed these quantities
by looking at the dependence of the thermodynamic variables on their
conjugate ones, that is, for example, the relation between entropy and
temperature, chemical potential and particle concentration, double
occupancy and on-site Coulomb repulsion. Once the self-consistent
equations for the single-particle propagator have been solved, we have
determined the temperature derivatives of the internal parameters by
means of exact linear systems of algebraic equations. The determination
of the first and second temperature derivatives of the chemical
potential has been revealed crucial in determining the thermodynamic
response functions under investigation. For the electronic specific heat
and internal energy we have presented three different schemes of
calculation. All of them allowed the possibility to obtain a deep
theoretical understanding of how and to which extent collective
excitations can be retained in the description of thermal response
functions. We have obtained a good agreement with the data by quantum
Monte Carlo techniques for the electronic specific heat and the internal
energy [34]. Further on, although Monte Carlo data shared common
features with the results from the calculations through the T-derivative
of the chemical potential, the experimental data for cuprates, as
revealed by the Wilson ratio and linear coefficient of the electronic
specific heat, have shown that in such systems the dominant excitations
are those of conventional non-interacting fermions [38].

We obtained several characteristic crossing points for the response
functions when reported as functions of some thermodynamic variables.
These peculiar features, already evidenced by Vollhardt [42], marked
turning points where different response functions evolve from a
non-interacting behaviour

\noindent (i) the entropy is an increasing function of U;

\noindent (ii) the entropy is an increasing function of $n$;

\noindent (iii) the double occupancy is a decreasing function of T;

\noindent (iv) the T-derivative of the chemical potential is a
decreasing function of $n$;

\noindent (v) the linear coefficient of the specific heat is an
increasing function of $n$;

to an unconventional dependence on the conjugate variables

\noindent (vi) the entropy is a decreasing function of U;

\noindent (vii) the entropy is a decreasing function of $n$;

\noindent (viii) the double occupancy is an increasing function of T;

\noindent (ix) the T-derivative of the chemical potential is an
increasing function of $n$;

\noindent (x) the linear coefficient of the specific heat is a
decreasing function of $n$.

 Before closing we would like to mention that the region of filling,
where (vi)-(x) hold, coincides with that where incommensurate magnetism
and superconductivity are experimentally observed in LSCO cuprates
family.

\acknowledgements

The authors thank Drs. D. Duffy and A. Moreo for kindly providing us
with data of the specific heat and internal energy. One of us (D.V.)
would like to thank Mr. Sarma Kancharla for a careful reading of the
manuscript. We also thank Mr. A. Avella and Prof. M. Marinaro for many
valuable discussions.
\newpage

\references

\item For a review, see, e.g., S. Uchida, Jpn. J. Appl. Phys. {\bf 32},
3784 (1993); Z.X. Shen and D.S. Dessau, Phys. Rep. {\bf 253}, 1 (1995).
\item E. Dagotto, Rev. Mod. Phys. {\bf 66}, 763 (1994).
\item P.W. Anderson, Science {\bf 235}, 1196 (1987).
\item A.P. Kampf, Phys. Rep. {\bf 249}, 219 (1994), and references
therein..
\item J. Hubbard, Proc. Roy. Soc. London, A {\bf 276}, 238 (1963).
\item Y. Tokura, Y. Taguchi, Y. Okada, Y. Fujishima, T. Arima, K.
Kumagai, Y. Iye, Phys. Rev. Lett. {\bf 70}, 2126 (1993).
\item K. Kumagai, T. Suzuki, Y. Taguchi, Y. Okada, Y. Fujishima, Y.
Tokura, Phys. Rev. B {\bf 48}, 7636 (1993).
\item D.B. McWhan, A. Menth, J.P. Remeika, W.F. Brinkman, T.M. Rice,
Phys. Rev. B {\bf 7}, 1920 (1973).
\item S.A. Carter, J. Yang, T.F. Rosenbaum, J. Spalek, J.M. Honig, Phys.
Rev. B {\bf 43}, 607 (1991).
\item S.A. Carter, T.F. Rosenbaum, J.M. Honig, J. Spalek, Phys. Rev.
Lett. {\bf 67}, 3440 (1991).
\item S.A. Carter, T.F. Rosenbaum, P. Metcalf, J.M. Honig, J. Spalek,
Phys. Rev. B {\bf 48}, 16841 (1991).
\item J.B. Torrance, A. Bezinge, A.I. Nazzal, T.C. Huang, S.S.P. Parkin,
D.T. Keane, S.J. LaPlaca, P.M. Horn, G.A. Held, Phys. Rev. B {\bf 40},
8872 (1989).
\item D.C. Johnston, Phys. Rev. Lett. {\bf 62}, 957 (1989).
\item S-W. Cheong, G. Aeppli, T.E. Mason, H. Mook, S.M. Hayden, P.C.
Canfield, Z. Fisk, K.N. Clausen, J.L. Martinez, Phys. Rev. Lett. {\bf
67}, 1791 (1991).
\item  G. Shirane, R.J. Birgeneau, Y. Endoh, M.A. Kastner, Physica B
{\bf 197}, 158 (1994).
\item  N. Ashcroft, N.D. Mermin, {\it Solid State Physics}, (Saunders,
1976).
\item  S. Marra, F. Mancini, A.M. Allega, H. Matsumoto, Physica C {\bf
235-240}, 2253 (1994).
\item F. Mancini, S. Marra, D. Villani, Condens. Matt. Phys. {\bf 7},
133 (1996).
\item A. Georges, W. Krauth, Phys. Rev. B {\bf 48}, 7167 (1993).
\item H. Matsumoto, M. Sasaki, I. Ishihara, M. Tachiki, Phys. Rev. B
{\bf 46}, 3009 (1992).
\item  H. Matsumoto, M. Sasaki, M. Tachiki, Phys. Rev. B {\bf 46}, 3022
(1992).
\item H. Matsumoto, S. Odashima, I. Ishihara, M. Tachiki, F. Mancini,
Phys. Rev. B {\bf 49}, 1350 (1994).
\item  F. Mancini, S. Marra, H. Matsumoto, Physica C {\bf 244}, 49
(1995).
\item F. Mancini, S. Marra, H. Matsumoto, Physica C {\bf 250}, 184
(1995).
\item F. Mancini, S. Marra, H. Matsumoto, Physica C {\bf 252}, 361
(1995).
\item  F. Mancini, H. Matsumoto, D. Villani, Czechoslovak Journ. of
Physics, {\bf 46} suppl. S4, 1871 (1996).
\item F. Mancini, S. Marra, H. Matsumoto, D. Villani, Phys. Lett. A {\bf
210}, 429 (1996)
\item F. Mancini, S. Marra, H. Matsumoto, Physica C {\bf 263}, 66
(1996).
\item F. Mancini, S. Marra, H. Matsumoto, Physica C {\bf 263}, 70
(1996).
\item F. Mancini, H. Matsumoto, D. Villani, Czechoslovak Journ. of
Physics, {\bf 46} suppl. S4, 1873 (1996).
\item H. Matsumoto, T. Saikawa, F. Mancini, Phys. Rev. B {\bf 54}, 14445
(1996).
\item H. Matsumoto, F. Mancini, Phys. Rev. B {\bf 55}, 2095 (1997).
\item F. Mancini, D. Villani, H. Matsumoto, Phys. Rev. B {\bf 57}, 6145
(1998).
\item D. Duffy, A. Moreo, cond-mat/9612132, 15 Dec 1996.
\item A. Avella, F. Mancini, D. Villani, L. Siurakshina and V. Yu.
Yushankhai, {\it The Hubbard model in the two-pole approximation},
cond-mat/9708009, Int. Journ. Mod. Phys. B (in print).
\item L.M. Roth, Phys. Rev. {\bf 184}, 451 (1969);
 W. Nolting and W. Borgel, Phys. Rev. B {\bf 39}, 6962 (1989); B.
Mehlig, H. Eskes, R. Hayn and M.B.J. Meinders, Phys. Rev. B {\bf 52},
2463 (1995).
\item J.W. Loram, K.A. Mirza, W.Y. Liang, J. Osborne, Physica C {\bf
162}, 498 (1989).
\item J.W. Loram, K.A. Mirza, J.R. Cooper, N. Athanasspoulou, W.Y.
Liang, {\it Thermodynamic evidence on the superconducting and normal
state energy gaps in $La_{2-x} Sr_x CuO_4$}, Preprint 1996.
\item J.W. Loram, K.A. Mirza, J.R. Cooper, W.Y. Liang, Phys. Rev. Lett.
{\bf 71}, 1740 (1993).
\item N. Wada, T. Obana, Y. Nakamura and K. Kumagai, Physica B {\bf
165\&166}, 1341 (1990).
\item Y. Okajima, K. Yamaya, N. Yamada, M. Oda, M. Ido, in "Mechanisms
of Superconductivity" JJAP Series 7, Ed. by Y. Muto, pag. 103 (1992).
\item D. Vollhardt, Phys. Rev. Lett. {\bf 78}, 1307 (1997).
\item D.F. Brewer, J.G. Daunt, A.K. Sreedhar, Phys. Rev. {\bf 115}, 836
(1959).
\item D.S. Greywall, Phys. Rev. B {\bf 27}, 2747 (1983).
\item G.E. Brodale, R.A. Fisher, N.E. Phillips, J. Flouquet, Phys. Rev.
Lett. {\bf 56}, 390 (1986).
\item N.E. Phillips, R.A. Fisher, J. Flouquet, A.L. Giorgi, J.A. Olsen,
G.R. Stewart, J. Magn. Magn. Mat. {\bf 63-64}, 332 (1987).
\item  A. de Visser, J.C.P. Klaasse, M. van Sprang, J.J.M. Franse, A.
Menovsky, T.T.M. Palstra, J. Magn. Magn. Mat. {\bf 54-57}, 375 (1986).
\item F. Steglich, C. Geibel, K. Gloos, G. Olesch, C. Schank, C.
Wassilew, A. Loidl, A. Krimmel, G.R. Stewart, J. Low Temp. Phys. {\bf
95}, 3 (1994).
\item A. Germann, H.v. Lohneysen, Europhys. Lett. {\bf 9}, 367 (1989).
\item H.G. Schlager, A. Schroder, M. Welsch, H.v. Lohneysen, J. Low
Temp. Phys. {\bf 90}, 181 (1993).
\item J. Fischer, A. Schr\"oder, H.v. L\"ohneysen, W. Bauhofer and U.
Steigenberger, Phys. Rev. B {\bf 39}, 11775 (1989).
\item J.W. Loram, K.A. Mirza, J.M. Wade, J.R. Cooper, W.Y. Liang,
Physica C {\bf 235-240}, 134 (1994).
\item  J. Schulte, M.C. Bohm, Phys. Rev. B {\bf 53}, 15385 (1996).
\item T. Usuki, N. Kawakami, A. Okiji, Journal of the Phys. Soc. of
Japan 59, 1357 (1990).
\item F. Gebhard, A. Girndt, A.E. Ruckenstein, Phys. Rev. B {\bf 49},
10926 (1994).
\item N. Bulut, D.J. Scalapino, S.R. White, Phys. Rev. B {\bf 50}, 7215
(1994); N. Bulut, D.J. Scalapino, S.R. White, Phys. Rev. Lett. {\bf 73},
748 (1994).
\item D. Duffy, A. Moreo, Phys. Rev. B {\bf 52}, 15607 (1995).
\item D.M. Newns, P.C. Pattnaik, C.C. Tsuei, Phys. Rev. B {\bf 43}, 3075
(1991).
\item J. Beenen, D.M. Edwards, Phys. Rev. B {\bf 52}, 13636 (1995).
\item Z.X. Shen, D.S. Dessau, B.O. Wells, D.M. King, J. Phys. Chem.
Solids {\bf 54}, 1169 (1993); D.S. Dessau et al. Phys. Rev. Lett. {\bf
71}, 2781 (1993).
\item M.L. Horbach, H. Kaj\"uter, Phys. Rev. Lett. {\bf 73}, 1309
(1994); M.L. Horbach, H. Kaj\"uter, Int. Journ. Mod. Phys. B {\bf 9}, 106
(1995).
\item R.S. Markiewicz, Phys. Rev. Lett. {\bf 73}, 1310 (1994).
\item J. Jaklic, P. Prelovsek, Phys. Rev. Lett. {\bf 77}, 892 (1996).
\item A.S. Alexandrov, N.F. Mott, Rep. Prog. Phys. {\bf 57}, 1197
(1994).
\item J. Spalek, A. Datta, J.M. Honig, Phys. Rev. Lett. {\bf 59}, 728
(1987).
\item J. Spalek, W. Wojcik, Phys. Rev. B {\bf 37}, 1532 (1988).
\item J. Spalek, Phys. Rev. B {\bf 40}, 5180 (1989).
\item J. Spalek, K. Byczuk, J. Karbowski, Physica Scripta T{\bf 49}, 206
(1993).
\item W. Loram, K.A. Mirza, J.R. Cooper, W.Y. Liang, J.M. Wade, J. of
Superconductivity {\bf 7}, 243 (1994).
\item K. Yamada, C.H. Lee, K. Kurahashi, J. Wada, S. Wakimoto, S. Ueki,
H. Kimura, Y. Endoh, S. Hosoya, G. Shirane, R.J. Birgeneau, M. Greven,
M.A. Kastner, Y.J. Kim, Phys. Rev. B {\bf 57}, 6165 (1998).

\end{document}